\documentclass[12pt,a4paper]{article}
\usepackage[margin=1.7cm]{geometry}
\usepackage{float}
\usepackage{subfigure}

\usepackage[footnotesize,sf]{caption}

\usepackage{sectsty}
\usepackage[normalem]{ulem}
\allsectionsfont{\sffamily\underline}
\usepackage[centertags]{amsmath}
\allowdisplaybreaks[1]
\usepackage{amsbsy}
\usepackage{amsfonts}
\usepackage{amssymb}
\usepackage{eufrak}
\usepackage[dvips]{graphicx}
\usepackage{url}
\usepackage[square,comma,numbers]{natbib}
\usepackage{tocbibind}

\newcommand{\blue}{\relax}

\newcommand{\red}{\relax}

\newcommand{\bluscuro}{\relax}

\newcommand{\porpora}{\relax}

\newcommand{\verdescuro}{\relax}
\newcommand{\viola}{\relax}
\newcommand{\violascuro}{\relax}
\newcommand{\semcm}{cm}
\begin{document}
\sf
\bluscuro

\null
\vspace{4cm}
\null

\Huge

\begin{center}
\text{\red{\underline{\porpora{
ESSENTIAL SOLAR NEUTRINOS
}}}}
\\[0.5cm]
\textrm{\violascuro{C. Giunti \& M. Laveder}}
\\[2cm]
%
\text{\verdescuro{31 January 2003}}
\\[2cm]
\fbox{
\begin{tabular}{c}
\textbf{\red{Neutrino Unbound}}
\\
\url{http://www.to.infn.it/~giunti/NU}
\end{tabular}
}
\\[2cm]
\texttt{hep-ph/0301276}
\end{center}

\normalsize

\newpage
{ \small
\tableofcontents
}
\newpage

\section{Standard Solar Model}
\label{Standard Solar Model}
\begin{center}
%
%
Current Standard Solar Model (SSM): BP2000 \protect\cite{Bahcall:2000nu,Bahcall-WWW}
\\[0.5cm]
\text{\blue{
$pp$ and CNO cycles:
}}
\text{\viola{
$
4 \, p + 2 \, e^-
\to
{}^4\mathrm{He} + 2 \, \nu_e + 26.731 \, \mathrm{MeV}
$
}}
\end{center}
\begin{figure}[H]
\begin{center}
\includegraphics*[bb=79 295 560 773, width=\linewidth]{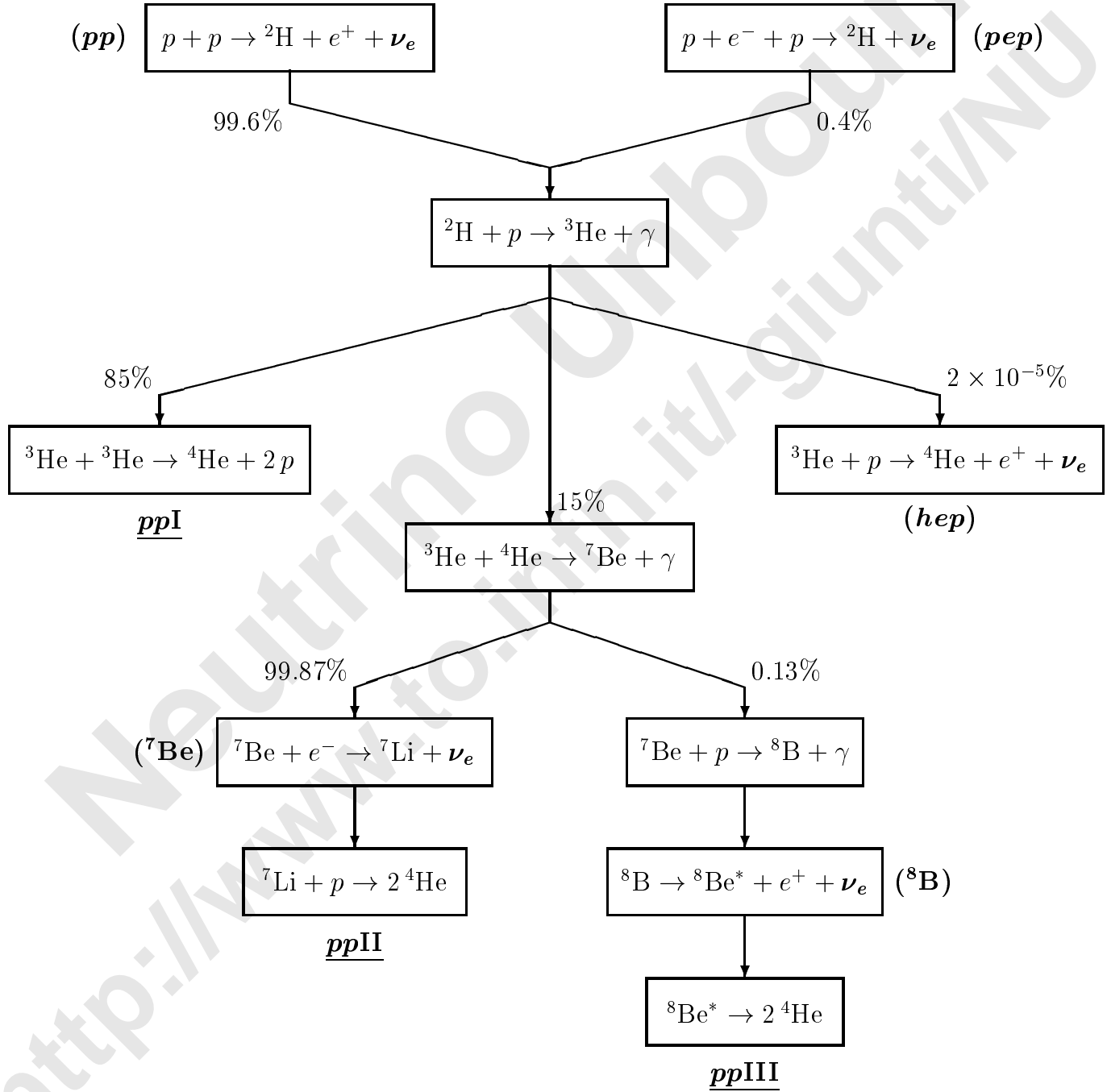}
\end{center}
\caption{ \label{sun_cycle_pp}
$pp$ cycle.
}
\end{figure}

\newpage

\begin{figure}[H]
\begin{center}
\includegraphics*[bb=100 512 473 773, width=0.7\linewidth]{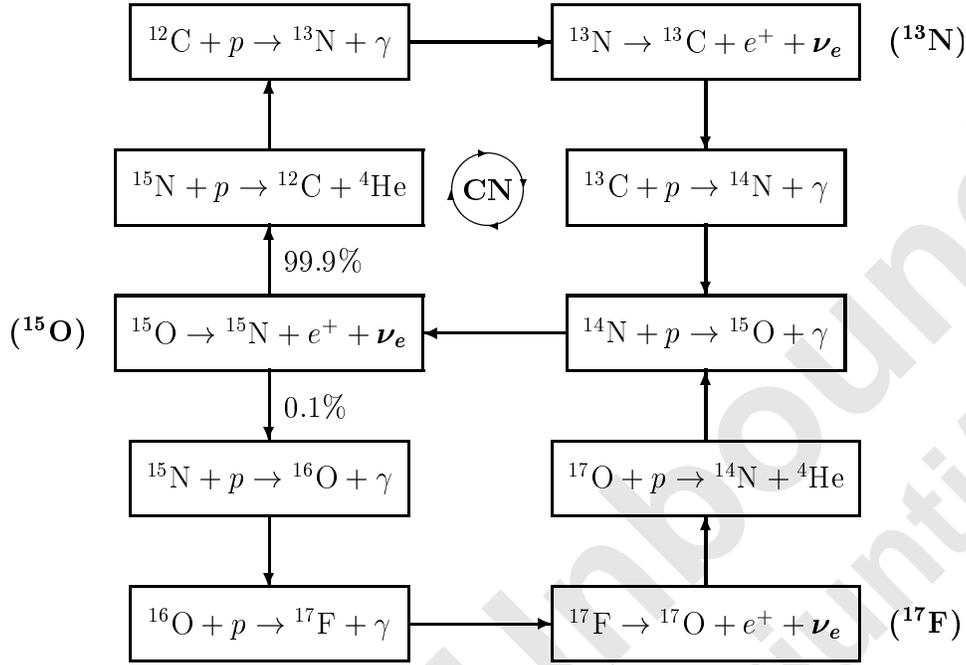}
\end{center}
\caption{ \label{sun_cycle_cno}
CNO cycle.
}
\end{figure}
\begin{table}[H]
\begin{center}
\begin{tabular}{|c|c|}
\hline
Luminosity
&
$ \mathcal{L}_\odot = ( 2.400 \pm 0.005 ) \times 10^{39} \, \mathrm{MeV} \, \mathrm{s}^{-1} $
\\
\hline
Radius
&
$ \mathcal{R}_\odot = 6.961 \times 10^{10} \, \mathrm{cm} $
\\
\hline
Mass
&
$ \mathcal{M}_\odot = ( 1.989 \pm 0.003 ) \times 10^{33} \, \mathrm{g} $
\\
\hline
Astronomical Unit
&
$ \mathrm{1a.u.} = 1.496 \times 10^{13} \, \mathrm{cm} $
\\
\hline
Solar Constant
&
$
K_\odot
\equiv
\mathcal{L}_\odot / 4 \pi (1\mathrm{a.u.})^2
=
8.534 \times 10^{11} \, \mathrm{MeV} \, \mathrm{cm}^{-2} \, \mathrm{s}^{-1}
$
\\
\hline
\end{tabular}
\end{center}
\caption{ \label{Sun}
Fundamental characteristics of the Sun
and Sun-Earth system
\protect\cite{PDG}.
One astronomical unit is the mean sun-earth distance.
The solar constant
$K_\odot$
is the mean solar photon flux on the Earth.
}
\end{table}

\begin{equation}
\text{Luminosity Constraint \text{\verdescuro{\protect\cite{Bahcall:2001pf}}}:}
\quad
\sum_r
\alpha_r
\,
\Phi_r
=
K_\odot
\qquad
(
r
=
pp,
pep,
hep,
{}^7\mathrm{Be},
{}^8\mathrm{B},
{}^{13}\mathrm{N},
{}^{15}\mathrm{O},
{}^{17}\mathrm{F}
)
\label{Luminosity}
\end{equation}

\section{Solar Neutrino Bibliography}
\label{Solar Neutrino Bibliography}
%
%
\begin{description}
\item[Books:]
\protect\cite{Rolfs-Rodney-book-88,Bahcall:1989ks}
\item[Reviews:]
\protect\cite{Castellani:1997cm,astro-ph/0210032,Miramonti-Reseghetti-2002}
\item[Bahcall's Standard Solar Models:]
\protect\cite{Bahcall:1964gx,Bahcall:1982zh,Bahcall:1988jc,Bahcall:1992hn,%
Bahcall:1995bt,Bahcall:1998wm,Bahcall:2000nu,Bahcall-WWW}
\item[Detection cross  sections:]
\protect\cite{Bahcall:1989ks,Bahcall-8B-96,Bahcall-Ga-97,Bahcall-WWW,%
Passera:2000ug,Butler:2000zp,Nakamura:2000vp,Ando:2002pv}
\end{description}

\newpage

\begin{figure}[H]
\begin{center}
\subfigure[
Figure from Ref.~\protect\cite{Castellani:1997cm}.
]{
\rotatebox{90}{
\includegraphics*[bb=102 80 550 746, height=0.475\linewidth]{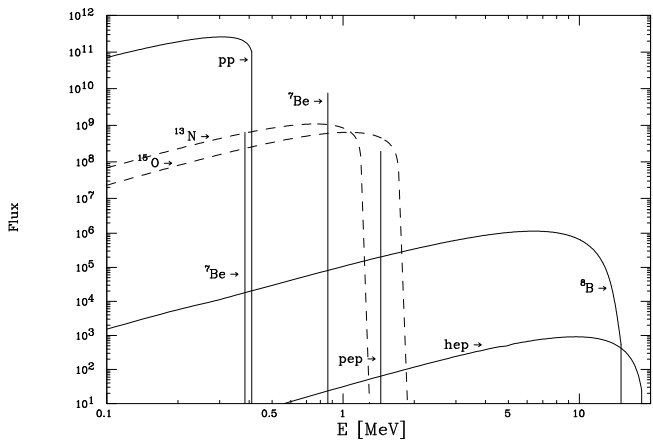}
}
\label{castellani-9606180-f04}
}
\hfill
\subfigure[
Figure from Ref.~\protect\cite{Bahcall-WWW}.
]{
\rotatebox[origin=rb]{270}{
\includegraphics*[bb=59 46 591 702, height=0.475\linewidth]{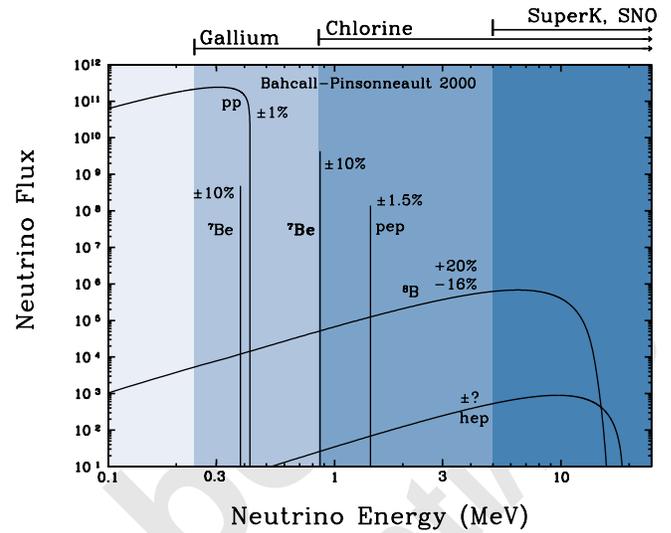}
}
\label{bahcall-spectra}
}
\end{center}
\caption{ \label{spectra}
Energy spectra of neutrino fluxes from the $pp$ and CNO chains,
as predicted by the Standard Solar Model.
For continuous sources, the differential flux is in
$\mathrm{cm}^{-2} \, \mathrm{s}^{-1} \, \mathrm{MeV}^{-1}$.
For the lines, the flux is in
$\mathrm{cm}^{-2} \, \mathrm{s}^{-1}$.
The percentages in Fig.~\ref{bahcall-spectra} indicate the uncertainties on the
values of the fluxes.
}
\end{figure}
\begin{figure}[H]
\begin{center}
\rotatebox{90}{
\includegraphics*[bb=92 106 570 685, height=0.60\linewidth]{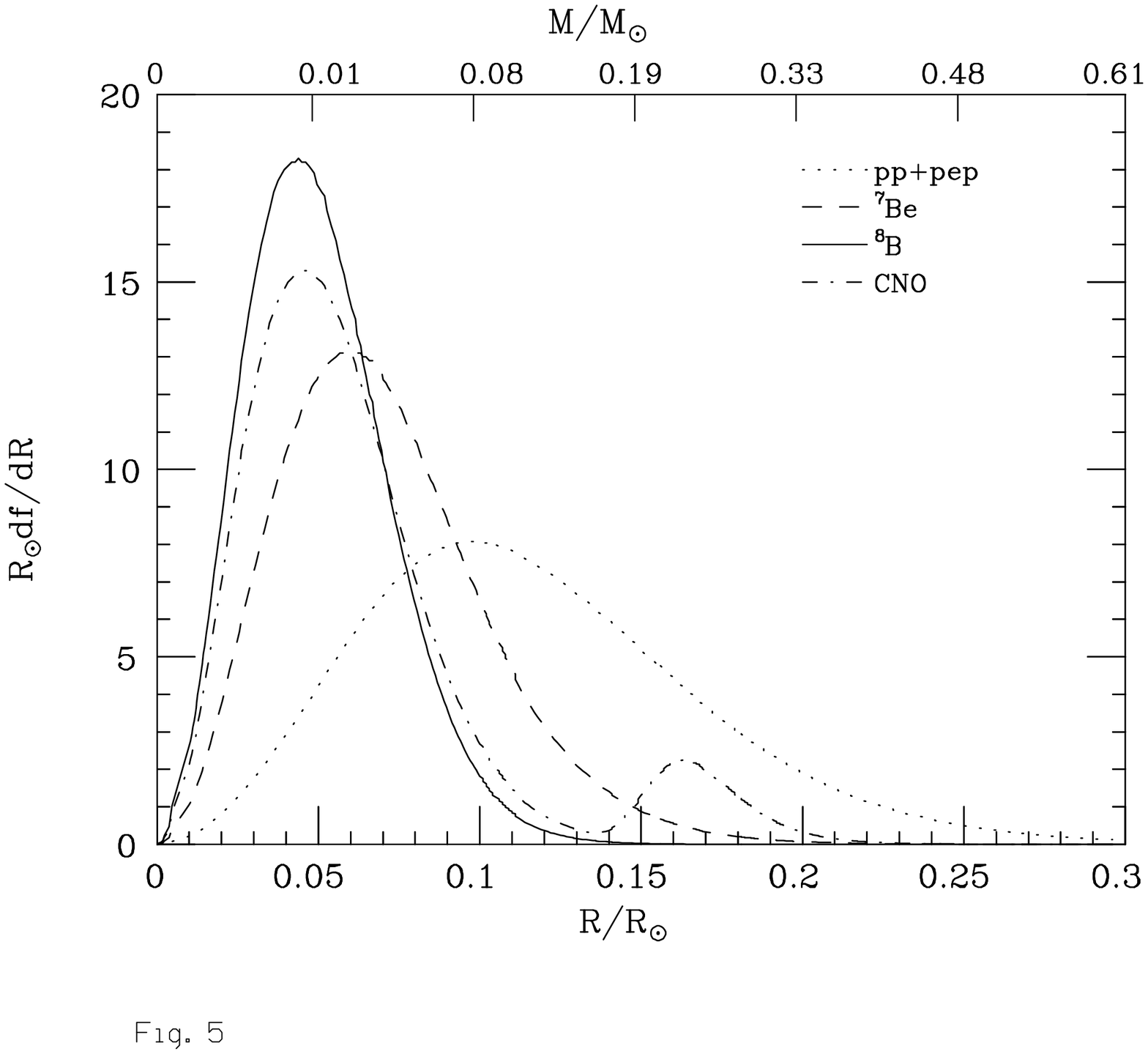}
}
\end{center}
\caption{ \label{castellani-9606180-f05}
Differential fraction $\mathrm{d}f/\mathrm{d}R$ of produced neutrinos as a function of radius $R$,
normalized to the solar radius $R_\odot$.
Figure from Ref.~\protect\cite{Castellani:1997cm}.
}
\end{figure}

\newpage

\begin{table}[H]
\begin{center}
\begin{tabular}{|c|c|c|c|c|}
\hline
Source $r$
&
Reaction
&
\begin{tabular}{c}
$\langle{E}\rangle_r$
\\
(MeV)
\end{tabular}
&
\begin{tabular}{c}
$E_r^{\mathrm{max}}$
\\
(MeV)
\end{tabular}
&
\begin{tabular}{c}
$\alpha_r$
\\
(MeV)
\end{tabular}
\\
\hline
\hline
$pp$
&
$ p + p \to d + e^+ + \nu_e $
&
$ 0.2668 $
&
$ 0.423 \pm 0.03 $
&
$ 13.0987 $
\vphantom{$\Big|$}
\\
\hline
$pep$
&
$ p + e^- + p \to d + \nu_e $
&
$ 1.445 $
&
$ 1.445 $
&
$ 11.9193 $
\vphantom{$\Big|$}
\\
\hline
$hep$
&
$ {}^3\mathrm{He} + p \to {}^4\mathrm{He} + e^+ + \nu_e $
&
$ 9.628 $
&
$ 18.778 $
&
$ 3.7370 $
\vphantom{$\Big|$}
\\
\hline
${}^7$Be
&
$ e^- + {}^7\mathrm{Be} \to {}^7\mathrm{Li} + \nu_e $
&
\begin{tabular}{c}
$ 0.3855 $
\\
$ 0.8631 $
\end{tabular}
&
\begin{tabular}{c}
$ 0.3855 $
\\
$ 0.8631 $
\end{tabular}
&
$ 12.6008 $
\vphantom{$\Big|$}
\\
\hline
${}^8$B
&
$ {}^8\mathrm{B} \to {}^8\mathrm{Be}^* + e^+ + \nu_e $
&
$ 6.735 \pm 0.036 $
&
$ \sim 15 $
&
$ 6.6305 $
\vphantom{$\Big|$}
\\
\hline
${}^{13}$N
&
$ {}^{13}\mathrm{N} \to {}^{13}\mathrm{C} + e^+ + \nu_e $
&
$ 0.7063 $
&
$ 1.1982 \pm 0.0003 $
&
$ 3.4577 $
\vphantom{$\Big|$}
\\
\hline
${}^{15}$O
&
$ {}^{15}\mathrm{O} \to {}^{15}\mathrm{N} + e^+ + \nu_e $
&
$ 0.9964 $
&
$ 1.7317 \pm 0.0005 $
&
$ 21.5706 $
\vphantom{$\Big|$}
\\
\hline
${}^{17}$F
&
$ {}^{17}\mathrm{F} \to {}^{17}\mathrm{O} + e^+ + \nu_e $
&
$ 0.9977 $
&
$ 1.7364 \pm 0.0003 $
&
$ 2.363 $
\vphantom{$\Big|$}
\\
\hline
\end{tabular}
\end{center}
\caption{ \label{Sources}
Sources of solar neutrinos
\protect\cite{Bahcall:1989ks,Bahcall-8B-96,Bahcall-Ga-97,Bahcall-Be-94}.
For each reaction $r$,
$\langle{E}\rangle_r$
is the average neutrino energy,
$E_r^{\mathrm{max}}$
is the maximum neutrino energy
and
$\alpha_r$
is the average thermal energy 
released together
with a neutrino from the source $r$
\protect\cite{Bahcall:2001pf},
that enters in the luminosity constraint (\ref{Luminosity}).
}
\end{table}

\begin{table}[H]
\begin{center}
\begin{tabular}{|c|c|c|c|c|c|}
\hline
Source $r$
&
\begin{tabular}{c}
Flux $\Phi_r$
\\
($ \mathrm{cm}^{-2} \, \mathrm{s}^{-1} $)
\end{tabular}
&
\begin{tabular}{c}
$\langle\sigma_{\mathrm{Cl}}\rangle_r$
\\
($ 10^{-44} \, \mathrm{cm}^2 $)
\end{tabular}
&
\begin{tabular}{c}
$S_{\mathrm{Cl}}^{(r)}$
\\
(SNU)
\end{tabular}
&
\begin{tabular}{c}
$\langle\sigma_{\mathrm{Ga}}\rangle_r$
\\
($ 10^{-44} \, \mathrm{cm}^2 $)
\end{tabular}
&
\begin{tabular}{c}
$S_{\mathrm{Ga}}^{(r)}$
\\
(SNU)
\end{tabular}
\\
\hline
\hline
$pp$
&
$ 5.95 \times 10^{10} \left( 1 \pm 0.01 \right) $
&
--
&
--
&
$ 0.117 \pm 0.003 $
&
$ 69.7 $
\vphantom{$\Big|$}
\\
\hline
$pep$
&
$ 1.40 \times 10^{8} \left( 1 \pm 0.015 \right) $
&
$ 0.16 $
&
$ 0.22 $
&
$ 2.04 \, {}^{+0.35}_{-0.14} $
&
$ 2.8 $
\vphantom{$\Big|$}
\\
\hline
$hep$
&
$ 9.3 \times 10^{3} $
&
$ 390 $
&
$ 0.04 $
&
$ 714 \, {}^{+228}_{-114} $
&
$ 0.1 $
\vphantom{$\Big|$}
\\
\hline
${}^7$Be
&
$ 4.77 \times 10^{9} \left( 1 \pm 0.10 \right) $
&
$ 0.024 $
&
$ 1.15 $
&
$ 0.717 \, {}^{+0.050}_{-0.021} $
&
$ 34.2 $
\vphantom{$\Big|$}
\\
\hline
${}^8$B
&
$ 5.05 \times 10^{6} \left( 1 \, {}^{+0.20}_{-0.16} \right) $
&
$ 114 \pm 11 $
&
$ 5.76 $
&
$ 240 \, {}^{+77}_{-36} $
&
$ 12.1 $
\vphantom{$\Big|$}
\\
\hline
${}^{13}$N
&
$ 5.48 \times 10^{8} \left( 1 \, {}^{+0.21}_{-0.17} \right) $
&
$ 0.017 $
&
$ 0.09 $
&
$ 0.604 \, {}^{+0.036}_{-0.018} $
&
$ 3.4 $
\vphantom{$\Big|$}
\\
\hline
${}^{15}$O
&
$ 4.80 \times 10^{8} \left( 1 \, {}^{+0.25}_{-0.19} \right) $
&
$ 0.068 \pm 0.001 $
&
$ 0.33 $
&
$ 1.137 \, {}^{+0.136}_{-0.057} $
&
$ 5.5 $
\vphantom{$\Big|$}
\\
\hline
${}^{17}$F
&
$ 5.63 \times 10^{6} \left( 1 \pm 0.25 \right) $
&
$ 0.069 $
&
$ 0.0 $
&
$ 1.139 \, {}^{+0.137}_{-0.057} $
&
$ 0.1 $
\vphantom{$\Big|$}
\\
\hline
\hline
Total
&
$ 6.54 \times 10^{10} $
&
&
$ 7.6 \, {}^{+1.3}_{-1.1} $
&
&
$ 128 \, {}^{+9}_{-7} $
\vphantom{$\Big|$}
\\
\hline
\end{tabular}
\end{center}
\caption{ \label{Fluxes}
BP2000 Standard Solar Model \protect\cite{Bahcall:2000nu}
neutrino fluxes,
average neutrino cross sections
\protect\cite{Bahcall:1989ks,Bahcall-8B-96,Bahcall-Ga-97}
and BP2000 SSM predictions
for the neutrino capture rates \protect\cite{Bahcall:2000nu}
in the chlorine (Cl) Homestake experiment
and in the gallium (Ga) GALLEX, SAGE and GNO experiments.
}
\end{table}

\newpage

\section{Homestake}
\label{Homestake}
\begin{center}
%
%
radiochemical experiment
\protect\cite{Bahcall:1964gx,Davis:1964hf}
\end{center}
\begin{equation}
\nu_e + ^{37}\mathrm{Cl}
\to
^{37}\mathrm{Ar} + e^-
\qquad
\text{\protect\cite{Pontecorvo-cl-46,Alvarez-cl-49}}
\label{cl-ar}
\end{equation}
\begin{center}
\text{\bluscuro{
Homestake Gold Mine (Lead, South Dakota, USA)
}}
\\[0.5cm]
\text{\bluscuro{
1478 m deep,
4200 m.w.e.
${\red{\Longrightarrow}}$
$\Phi_\mu \simeq 4 \, \mathrm{m}^{-2} \, \mathrm{day}^{-1}$
}}
\\[0.5cm]
\text{\bluscuro{
steel tank, 6.1 m diameter, 14.6 m long
($6 \times 10^5$ liters)
}}
\\[0.5cm]
\text{\bluscuro{
615 tons of tetrachloroethylene ($\mathrm{C}_2\mathrm{Cl}_4$),
$2.16 \times 10^{30}$ atoms of $^{37}\mathrm{Cl}$
(133 tons)
}}
\\[0.5cm]
\text{\violascuro{
energy threshold:
$ E_{\mathrm{th}}^{\mathrm{Cl}} = 0.814 \, \mathrm{MeV} $
${\red{\Longrightarrow}}$
$^8\mathrm{B}$,
$^7\mathrm{Be}$,
$pep$,
$hep$
}}
\\[0.5cm]
\text{
data taking: 1970--1994, 108 extractions
\protect\cite{Cleveland:1998nv}
--
history: \protect\cite{Bahcall:2000up,Bahcall:2000rt}
}
\end{center}
\begin{figure}[H]
\begin{center}
\includegraphics*[bb=325 109 560 258, height=0.35\textheight]{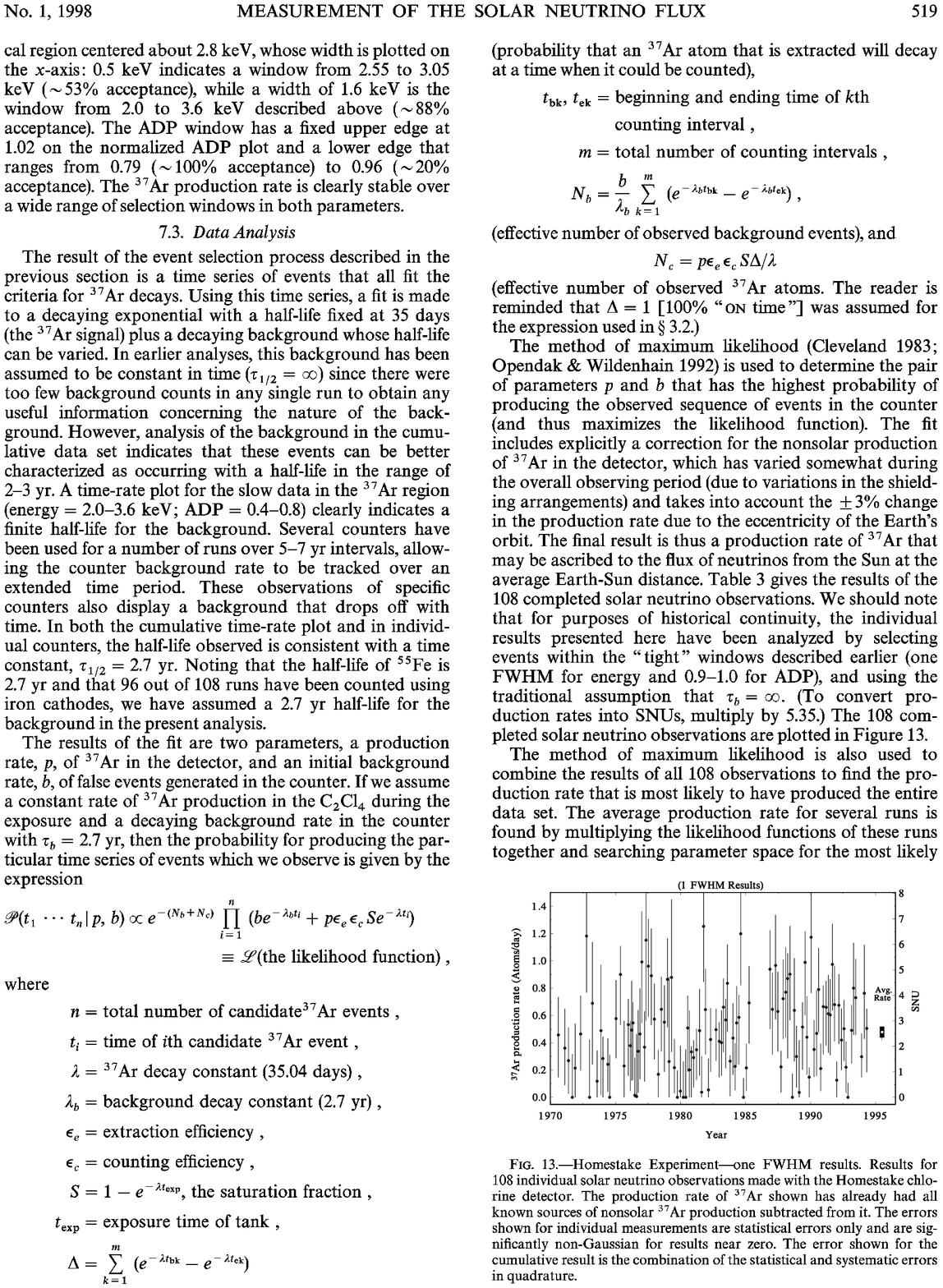}
\end{center}
\caption{ \label{homestake-f13}
Results of the 108 individual solar neutrino
observations made with the Homestake chlorine detector.
The production rate of $^{37}\mathrm{Ar}$ shown has
already had all known sources of nonsolar $^{37}\mathrm{Ar}$ production subtracted from it.
The errors shown for
individual measurements are statistical errors only and are significantly non-Gaussian for results
near zero. The error shown for the cumulative result is the combination of the statistical and
systematic errors in quadrature.
Figure from Ref.~\protect\cite{Cleveland:1998nv}.
}
\end{figure}
\begin{equation}
R_{\mathrm{Cl}}^{\mathrm{exp}}
=
2.56 \pm 0.16 \pm 0.16 \, \mathrm{SNU}
=
2.56 \pm 0.23 \, \mathrm{SNU}
\qquad
\text{\protect\cite{Cleveland:1998nv}}
\label{Rclexp}
\end{equation}
\begin{equation}
R_{\mathrm{Cl}}^{\mathrm{SSM}}
=
7.6 {}^{+1.3}_{-1.1} \, \mathrm{SNU}
\qquad
\text{\protect\cite{Bahcall:2000nu,Bahcall-WWW}}
\label{Rclssm}
\end{equation}

\newpage

\section{Gallium Experiments: SAGE, GALLEX, GNO}
\label{Gallium Experiments: SAGE, GALLEX, GNO}
\begin{center}
%
%
radiochemical experiments
\end{center}
\begin{equation}
\nu_e + {}^{71}\mathrm{Ga} \to {}^{71}\mathrm{Ge} + e^-
\qquad
\text{\protect\cite{Kuzmin-Ga-65}}
\label{ga-ge}
\end{equation}
\begin{center}
\text{\violascuro{
threshold:
$ E_{\mathrm{th}}^{\mathrm{Ga}} = 0.233 \, \mathrm{MeV} $
${\red{\Longrightarrow}}$
all $\nu$ fluxes
($pp$,
$^7\mathrm{Be}$,
$^8\mathrm{B}$,
$pep$,
$hep$,
$^{13}\mathrm{N}$,
$^{15}\mathrm{O}$,
$^{17}\mathrm{F}$)
}}
\end{center}
\begin{equation}
\text{SAGE + GALLEX + GNO}
\quad
\Longrightarrow
\quad
R_{\mathrm{Ga}}^{\mathrm{exp}}
=
72.4 \pm 4.7 \, \mathrm{SNU}
\label{Rgaexp}
\end{equation}
\begin{equation}
\text{Standard Solar Model}
\quad
\Longrightarrow
\quad
R_{\mathrm{Ga}}^{\mathrm{SSM}}
=
128 {}^{+9}_{-7} \, \mathrm{SNU}
\qquad
\text{\protect\cite{Bahcall:2000nu,Bahcall-WWW}}
\label{Rgassm}
\end{equation}

\section{SAGE: Soviet-American Gallium Experiment}
\label{SAGE: Soviet-American Gallium Experiment}
\begin{center}
%
%
\text{
Baksan Neutrino Observatory, northern Caucasus,
3.5 km from entrance of horizontal adit
}
\\[0.5\semcm]
\text{
50 tons of metallic $^{71}\mathrm{Ga}$,
2000 m deep,
4700 m.w.e.
${\red{\Longrightarrow}}$
$\Phi_\mu \simeq 2.6 \, \mathrm{m}^{-2}  \, \mathrm{day}^{-1} $ 
}
\\[0.5\semcm]
\text{
data taking: 1990 -- 2001, 92 runs
\protect\cite{Abazov:1991rx,%
Abdurashitov:1994bc,%
Abdurashitov:1999bv,%
Abdurashitov:1999zd,%
Abdurashitov:2002nt}
}
\\[0.5\semcm]
\text{
detector test:
$^{51}\mathrm{Cr}$ Source
($ R = 0.95 {}^{+0.11}_{-0.10} {}^{+0.06}_{-0.05} $)
\protect\cite{Abdurashitov:1996dp,Abdurashitov:1998ne}
}
\end{center}
\begin{figure}[H]
\begin{center}
\includegraphics*[bb=37 36 449 287, height=0.35\textheight]{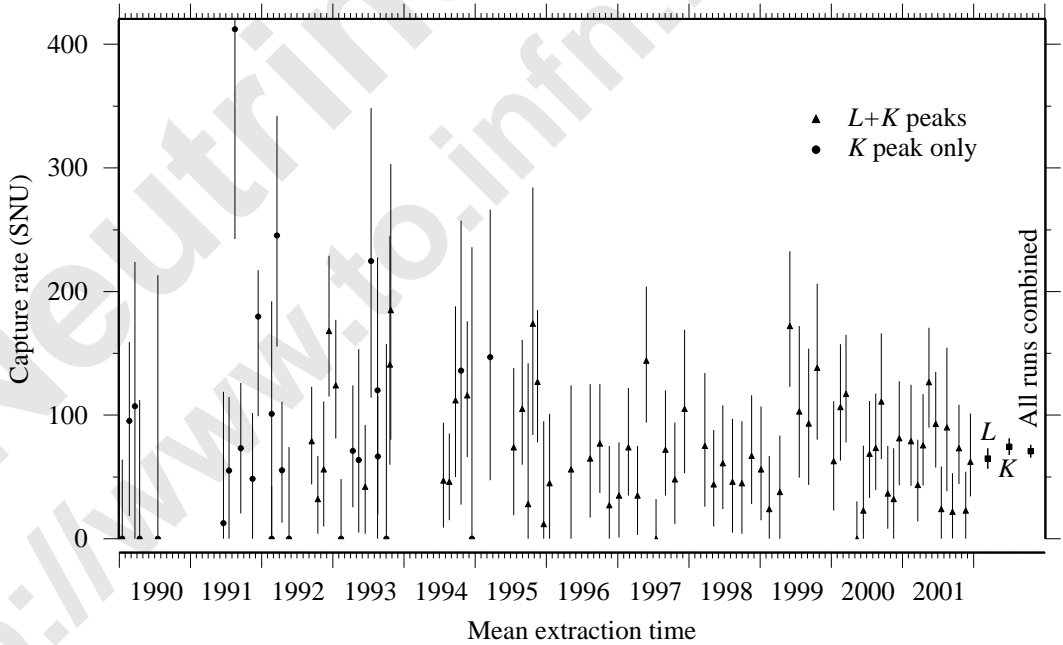}
\end{center}
\caption{ \label{sage-0204245-f01}
Capture rate for all SAGE extractions as a function of time.  
Error bars are statistical with 68\% confidence.
The combined result
of all runs in the $L$ peak, the $K$ peak, and both $L$ and $K$ peaks
is shown on the right side.
The last 3 runs are still counting and 
their results are preliminary.
Figure from Ref.~\protect\cite{Abdurashitov:2002nt}.
}
\end{figure}
\begin{equation}
R_{\mathrm{Ga}}^{\mathrm{SAGE}}
=
70.8 {}^{+5.3}_{-5.2} {}^{+3.7}_{-3.2} \, \mathrm{SNU}
=
70.8 {}^{+6.5}_{-6.1} \, \mathrm{SNU}
\qquad
\text{\protect\cite{Abdurashitov:2002nt}}
\label{Rsage}
\end{equation}

\newpage

\section{GALLEX: GALLium EXperiment}
\label{GALLEX: GALLium EXperiment}
\begin{center}
%
%
\text{
Gran Sasso Underground Laboratory, Italy,
overhead shielding: 3300 m.w.e.
}
\\[0.5\semcm]
\text{
30.3 tons of gallium
in 101 tons of gallium chloride
(GaCl$_3$-HCl)
solution
}
\\[0.5\semcm]
\text{
data taking: May 1991 -- Jan 1997, 65 runs
\protect\cite{Anselmann:1992um,%
Anselmann:1993mh,%
Anselmann:1994cf,%
Anselmann:1995ag,%
Hampel:1996qd,%
Hampel:1998xg}
}
\\[0.5\semcm]
\text{
detector tests:
$^{51}\mathrm{Cr}$ Source
($ R = 0.93 \pm 0.08 $)
\protect\cite{Anselmann:1995ar,Hampel:1998fc},
$^{71}\mathrm{As}$ Test
\protect\cite{Hampel:1998su}
}
\end{center}
\begin{equation}
R_{\mathrm{Ga}}^{\mathrm{GALLEX}}
=
77.5 \pm 6.2 {}^{+4.3}_{-4.7} \, \mathrm{SNU}
=
77.5 {}^{+7.6}_{-7.8} \, \mathrm{SNU}
\qquad
\text{\protect\cite{Hampel:1998xg}}
\label{Rgallex}
\end{equation}

\section{GNO: Gallium Neutrino Observatory}
\label{GNO: Gallium Neutrino Observatory}
\begin{center}
%
%
\text{
successor of GALLEX,
GNO30: 30.3 tons of gallium
}
\\[0.5\semcm]
\text{
data taking: May 1998 -- Jan 2000, 19 runs
\protect\cite{Altmann:2000ft}
}
\end{center}
\begin{equation}
R_{\mathrm{Ga}}^{\mathrm{GNO}}
=
65.8 {}^{+10.2}_{-9.6} {}^{+3.4}_{-3.6} \, \mathrm{SNU}
=
65.8 {}^{+10.7}_{-10.2} \, \mathrm{SNU}
\qquad
\text{\protect\cite{Altmann:2000ft}}
\label{Rgno}
\end{equation}
\begin{figure}[H]
\begin{center}
\includegraphics*[bb=19 131 767 471, height=0.35\textheight]{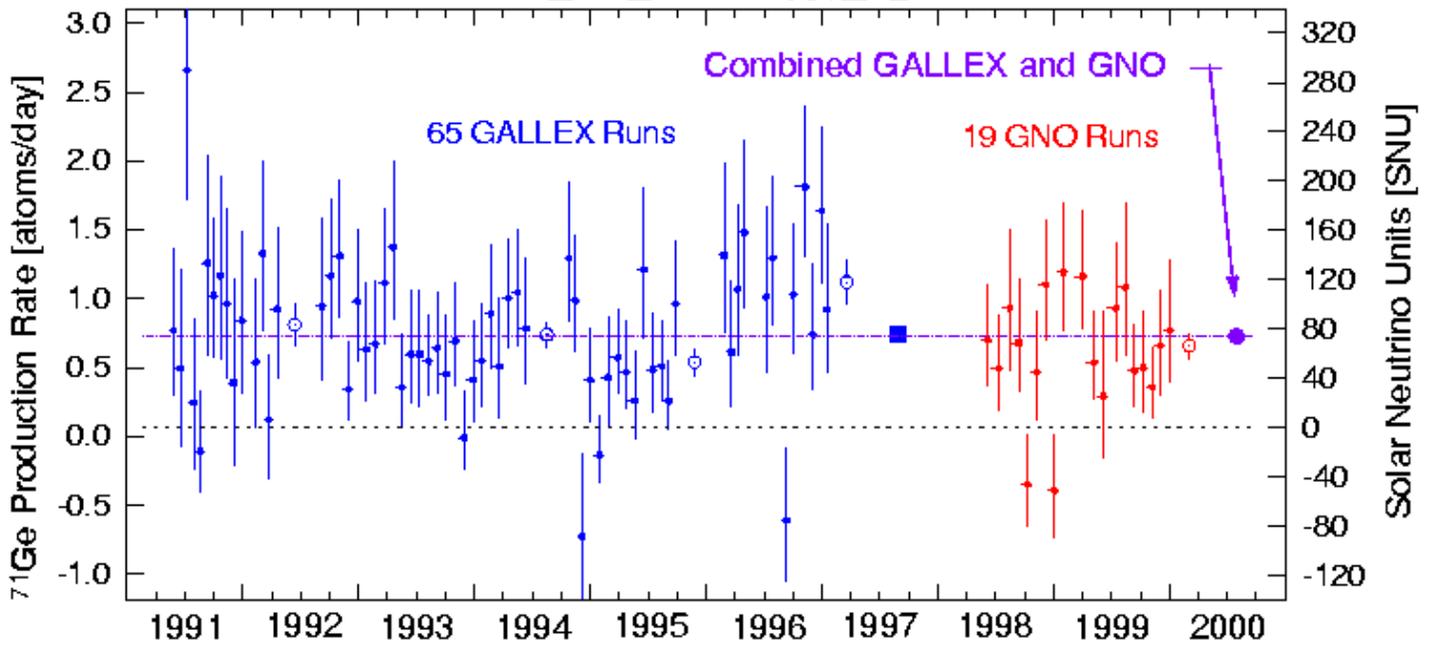}
\end{center}
\caption{ \label{gno-0006034-f03}
GNO and GALLEX single run results.
Error bars are statistical only.
Open dots represent group mean values;
the bold square represents the global result of GALLEX;
the bold solid dot represents the global result of GNO and Gallex.
Figure from Ref.~\protect\cite{Altmann:2000ft}.
}
\end{figure}
\begin{equation}
\text{GALLEX + GNO}
\quad
\Longrightarrow
\quad
R_{\mathrm{Ga}}^{\mathrm{GALLEX+GNO}}
=
74.1 \pm 5.4 {}^{+4.0}_{-4.2} \, \mathrm{SNU}
=
74.1 {}^{+6.7}_{-6.8} \, \mathrm{SNU}
\qquad
\text{\protect\cite{Altmann:2000ft}}
\label{Rgallexgno}
\end{equation}

\newpage

\section{Kamiokande}
\label{Kamiokande}
\begin{center}
%
%
\text{
real-time water Cherenkov detector
}
\hspace{1\semcm}
\text{\red{
$\nu + e^- \to \nu + e^-$
}}
\hspace{1\semcm}
\protect\cite{Koshiba:1992yb,Totsuka:1992dm}
\\[0.5\semcm]
\text{\violascuro{
Sensitive to
$\nu_e$,
$\nu_\mu$,
$\nu_\tau$,
but
$
\sigma(\nu_e) \simeq 6 \, \sigma(\nu_{\mu,\tau})
$
}}
\\[0.5\semcm]
\text{
Kamioka mine (200 km west of Tokyo),
1000 m underground,
2700 m.w.e.
}
\\[0.5\semcm]
\text{
3000 tons of water,
680 tons fiducial volume,
948 PMTs
}
\\[0.5\semcm]
\text{\violascuro{
threshold:
$ E_{\mathrm{th}}^{\mathrm{Kam}} \simeq 6.75 \, \mathrm{MeV} $
${\red{\Longrightarrow}}$
$^8\mathrm{B}$,
$hep$
}}
\\[0.5\semcm]
\text{
data taking: Jan 1987 -- Feb 1995 (2079 days)
\protect\cite{Hirata:1989zj,%
Hirata:1990fj,%
Hirata:1990xa,%
Hirata:1991ep,%
Hirata:1991ub,%
Fukuda:1996sz}
}
\end{center}
\begin{equation}
\Phi_{{\nu}e}^{\mathrm{Kam}}
=
2.82 {}^{+0.25}_{-0.24} \pm 0.27 \times 10^{6} \, \mathrm{cm}^{-2} \mathrm{s}^{-1}
=
2.82 \pm 0.37 \times 10^{6} \, \mathrm{cm}^{-2} \mathrm{s}^{-1}
\qquad
\text{\protect\cite{Fukuda:1996sz}}
\label{PhiKam}
\end{equation}
\begin{equation}
\text{Standard Solar Model}
\quad
\Longrightarrow
\quad
\Phi_{{\nu}e}^{^8\mathrm{B}}
=
5.05 {}^{+1.01}_{-0.81} \times 10^{6} \, \mathrm{cm}^{-2} \mathrm{s}^{-1}
\qquad
\text{\protect\cite{Bahcall:2000nu,Bahcall-WWW}}
\label{PhiSSM}
\end{equation}

\section{Super-Kamiokande}
\label{Super-Kamiokande}
\begin{center}
%
%
\text{
successor of Kamiokande,
50 ktons of water,
22.5 ktons fiducial volume,
11146 PMTs
}
\\[0.5\semcm]
\text{\violascuro{
threshold:
$ E_{\mathrm{th}}^{\mathrm{Kam}} \simeq 4.75 \, \mathrm{MeV} $
${\red{\Longrightarrow}}$
$^8\mathrm{B}$,
$hep$
}}
\\[0.5\semcm]
\text{
data taking: 1996 -- 2001 (1496 days)
\protect\cite{Fukuda:1998fd,%
Fukuda:1998rq,%
Fukuda:1998ua,%
Fukuda:2001nj,%
Fukuda:2001nk,%
Fukuda:2002pe}
}
\end{center}
\begin{equation}
\Phi_{{\nu}e}^{\mathrm{SK}}
=
2.348 \pm 0.025 {}^{+0.071}_{-0.061} \times 10^{6} \, \mathrm{cm}^{-2} \mathrm{s}^{-1}
=
2.348 {}^{+0.075}_{-0.066} \times 10^{6} \, \mathrm{cm}^{-2} \mathrm{s}^{-1}
\qquad
\text{\protect\cite{Fukuda:2002pe}}
\label{PhiSK}
\end{equation}
\begin{figure}[H]
\begin{center}
\includegraphics*[bb=0 0 620 540, height=0.25\textheight]{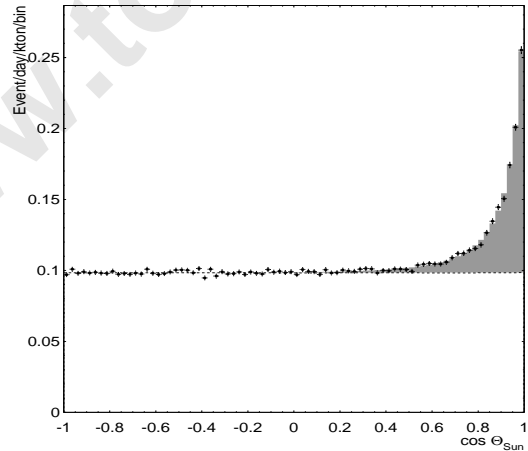}
\end{center}
\caption{ \label{sk-smy-0208004-f01a}
Super-Kamiokande $\cos\theta_{\mathrm{sun}}$ distribution.
The points represent observed data.
The histogram
shows the best-fit signal (shaded) plus background.
The horizontal dashed line shows the estimated background.
Figure from Ref.~\protect\cite{Smy:2002rz}.
}
\end{figure}

\newpage

\begin{figure}[H]
\begin{center}
\includegraphics*[bb=0 0 1239 883, height=0.21\textheight]{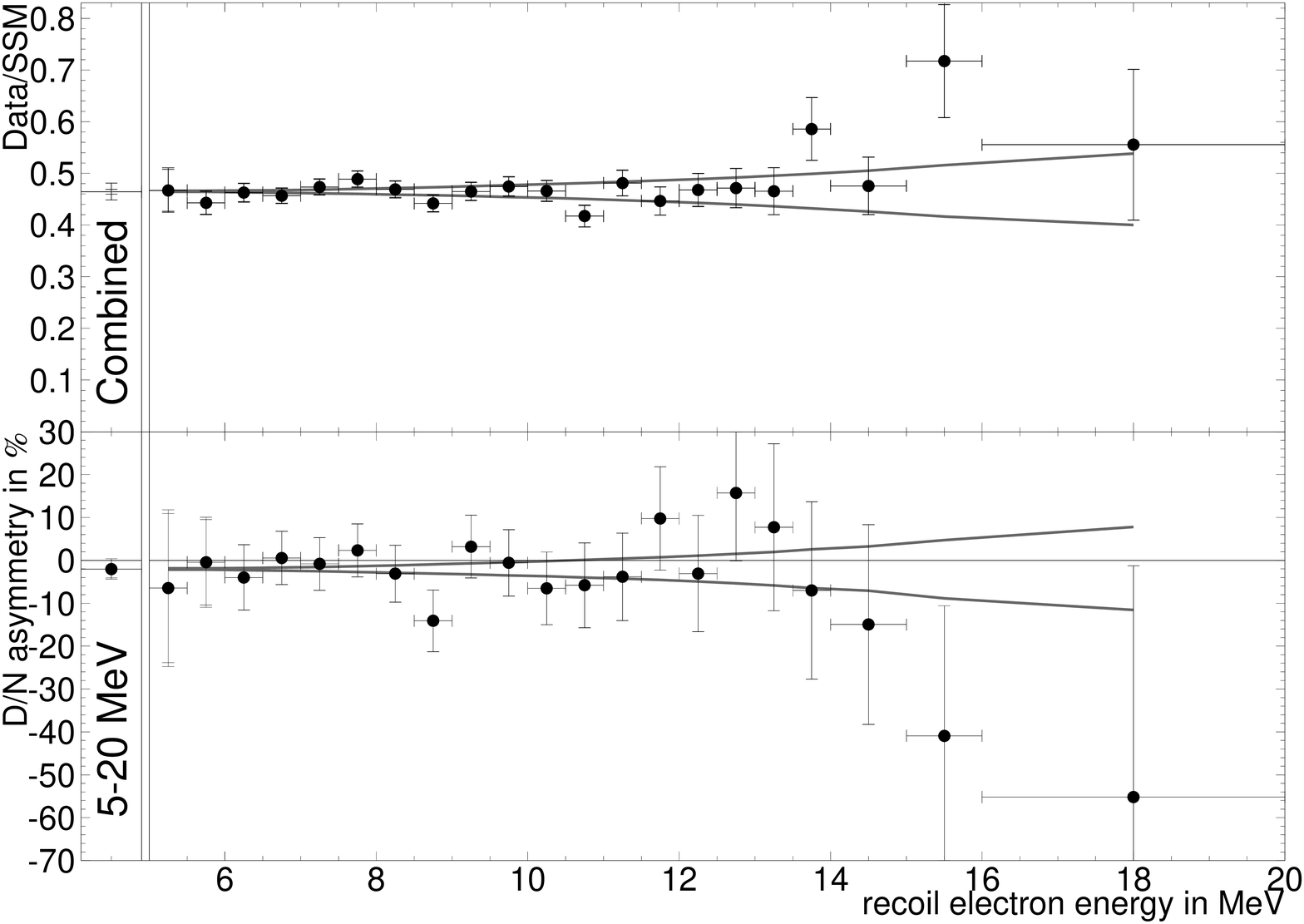}
\end{center}
\caption{ \label{sk-smy-0208004-f05}
Super-Kamiokande energy spectrum
normalized to BP2000 SSM \protect\cite{Smy:2002rz}.
}
\end{figure}
\begin{figure}[H]
\begin{center}
\includegraphics*[bb=0 0 1258 839, height=0.21\textheight]{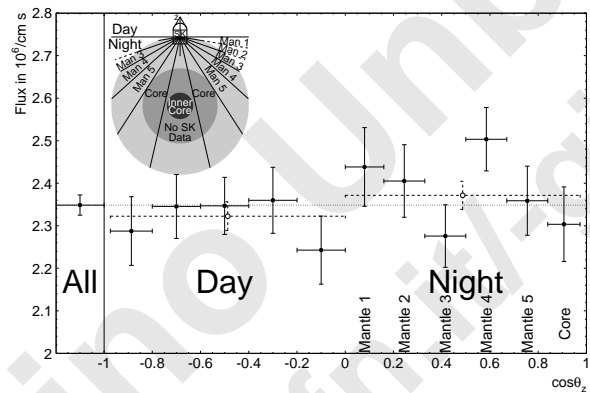}
\end{center}
\caption{ \label{sk-smy-0208004-f04}
Solar zenith angle ($\theta_z$) dependence
of Super-Kamiokande data \protect\cite{Smy:2002rz}.
}
\end{figure}
\begin{figure}[H]
\begin{center}
\includegraphics*[bb=0 5 1121 515, height=0.27\textheight]{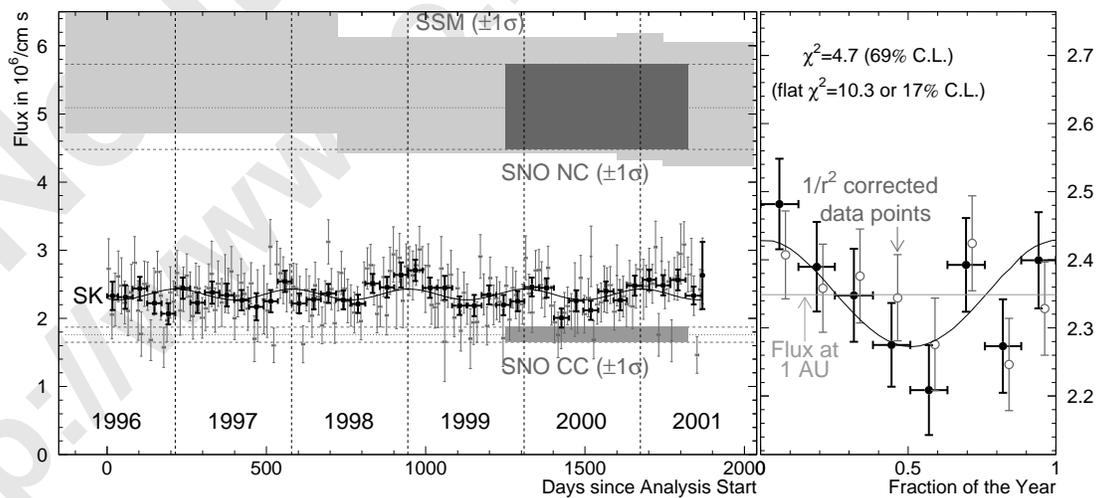}
\end{center}
\caption{ \label{sk-smy-0208004-f03}
Time variation of the Super-Kamiokande data.
The gray
data points are measured every 10 days, the black data points every
1.5 months.
The black line indicates the expected annual 7\% flux variation.
The right-hand panel combines the 1.5 month bins to search for
yearly variations.
The gray data points (open circles)
are obtained from the
black data points
by subtracting the expected 7\% variation.
Figure from Ref.~\protect\cite{Smy:2002rz}.
}
\end{figure}

\newpage

\section{SNO: Sudbury Neutrino Observatory}
\label{SNO: Sudbury Neutrino Observatory}
%
%
\begin{tabular}{lcr}
\begin{minipage}[t]{0.5\textwidth}
\begin{tabular}{c}
\text{
real-time heavy-water Cherenkov detector
}
\\[0.5\semcm]
\text{
Creighton mine (INCO Ltd.), Sudbury, Ontario, Canada
}
\\[0.5\semcm]
\text{
1 kton of $\mathrm{D}_2\mathrm{O}$, 9456 20-cm PMTs
}
\\[0.5\semcm]
\text{
2073 m underground,
6010 m.w.e.
}
\end{tabular}
\end{minipage}
& \hspace{0.1\textwidth} &
\begin{minipage}[t]{0.3\textwidth}
$
\begin{array}{ll}
\text{\red{CC:}}
&
\nu_e + d \to p + p + e^-
\\[0.5\semcm]
\text{\red{NC:}}
&
\nu + d \to p + n + \nu
\\[0.5\semcm]
\text{\red{ES:}}
&
\nu + e^- \to \nu + e^-
\end{array}
$
\end{minipage}
\end{tabular}
\begin{center}
\text{\red{
$
\left.
\begin{array}{l}
\text{\violascuro{
CC threshold:
$ E_{\mathrm{th}}^{\mathrm{SNO}}(\mathrm{CC}) \simeq 8.2 \, \mathrm{MeV} $
}}
\\[0.5\semcm]
\text{\violascuro{
NC threshold:
$ E_{\mathrm{th}}^{\mathrm{SNO}}(\mathrm{NC}) \simeq 2.2 \, \mathrm{MeV} $
}}
\\[0.5\semcm]
\text{\violascuro{
ES threshold:
$ E_{\mathrm{th}}^{\mathrm{SNO}}(\mathrm{ES}) \simeq 7.0 \, \mathrm{MeV} $
}}
\end{array}
\right\}
$
${\red{\Longrightarrow}}$
$^8\mathrm{B}$,
$hep$
}}
\\[0.5\semcm]
\text{
data taking: 1999 -- 2002 (306.4 days)
\protect\cite{Ahmad:2001an,Ahmad:2002jz,Ahmad:2002ka}
}
\end{center}
\begin{align}
\null & \null
\Phi_{\mathrm{CC}}^{\mathrm{SNO}}
=
1.76 {}^{+0.06}_{-0.05} \pm 0.09 \times 10^{6} \, \mathrm{cm}^{-2} \mathrm{s}^{-1}
=
1.76 {}^{+0.11}_{-0.10} \times 10^{6} \, \mathrm{cm}^{-2} \mathrm{s}^{-1}
\qquad
\text{\protect\cite{Ahmad:2002jz}}
\label{PhiSNOCC}
\\
\null & \null
\Phi_{\mathrm{NC}}^{\mathrm{SNO}}
=
5.09 {}^{+0.44}_{-0.43} {}^{+0.46}_{-0.43} \times 10^{6} \, \mathrm{cm}^{-2} \mathrm{s}^{-1}
=
5.09 {}^{+0.64}_{-0.61} \times 10^{6} \, \mathrm{cm}^{-2} \mathrm{s}^{-1}
\qquad
\text{\protect\cite{Ahmad:2002jz}}
\label{PhiSNONC}
\\
\null & \null
\Phi_{\mathrm{ES}}^{\mathrm{SNO}}
=
2.39 {}^{+0.24}_{-0.23} \pm 0.12 \times 10^{6} \, \mathrm{cm}^{-2} \mathrm{s}^{-1}
=
2.39 {}^{+0.27}_{-0.26} \times 10^{6} \, \mathrm{cm}^{-2} \mathrm{s}^{-1}
\qquad
\text{\protect\cite{Ahmad:2002jz}}
\label{PhiSNOES}
\end{align}
\begin{align}
\null & \null
\Phi_{\nu_e}^{\mathrm{SNO}}
=
1.76 \pm 0.05 \pm 0.09 \times 10^{6} \, \mathrm{cm}^{-2} \mathrm{s}^{-1}
=
1.76 \pm 0.10 \times 10^{6} \, \mathrm{cm}^{-2} \mathrm{s}^{-1}
\qquad
\text{\protect\cite{Ahmad:2002jz}}
\label{PhiSNOnue}
\\
\null & \null
\Phi_{\nu_{\mu,\tau}}^{\mathrm{SNO}}
=
3.41 \pm 0.45 {}^{+0.48}_{-0.45} \times 10^{6} \, \mathrm{cm}^{-2} \mathrm{s}^{-1}
=
3.41 {}^{+0.66}_{-0.64} \times 10^{6} \, \mathrm{cm}^{-2} \mathrm{s}^{-1}
\qquad
\text{\protect\cite{Ahmad:2002jz}}
\label{PhiSNOnum}
\end{align}
\begin{center}
$5.3\sigma$ evidence of $\nu_e\to\nu_{\mu,\tau}$ transitions
\end{center}
\begin{figure}[H]
\begin{center}
\includegraphics*[bb=7 0 520 390, height=0.29\textheight]{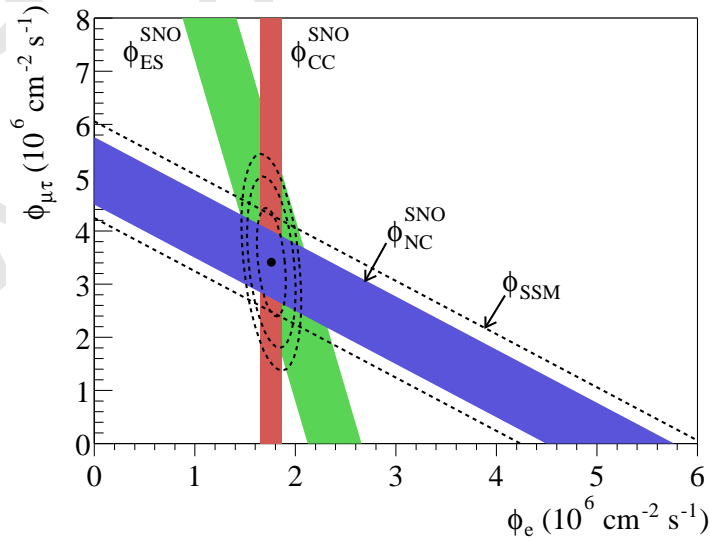}
\end{center}
\caption{ \label{sno-0204008-f03}
Flux of $\nu_\mu$ and $\nu_\tau$ vs flux of $\nu_e$
in the ${}^{8}$B energy range
deduced from the three neutrino reactions in SNO.
The diagonal bands show the total BP2000 ${}^{8}$B flux \protect\cite{Bahcall:2000nu} (dashed lines)
and that measured with the NC reaction in SNO (solid band).
The intercepts of these bands with the axes represent the $\pm 1\sigma$ errors.
The bands intersect at the fit values for
$\phi_{e} \equiv \Phi_{\nu_e}$ and $\phi_{\mu\tau} \equiv \Phi_{\nu_{\mu,\tau}}$.
Figure from Ref.~\protect\cite{Ahmad:2002jz}.
}
\end{figure}

\newpage

\begin{figure}[H]
\begin{center}
\subfigure[
Solar zenith angle ($\theta_\odot$) dependence
of SNO data.
ES: $ \cos\theta_\odot \simeq 1 $.
CC: $ \sigma \propto 1 - 0.340 \cos\theta_\odot $.
NC: isotropic.
]{
\includegraphics*[bb=6 491 316 731, width=0.45\textwidth]{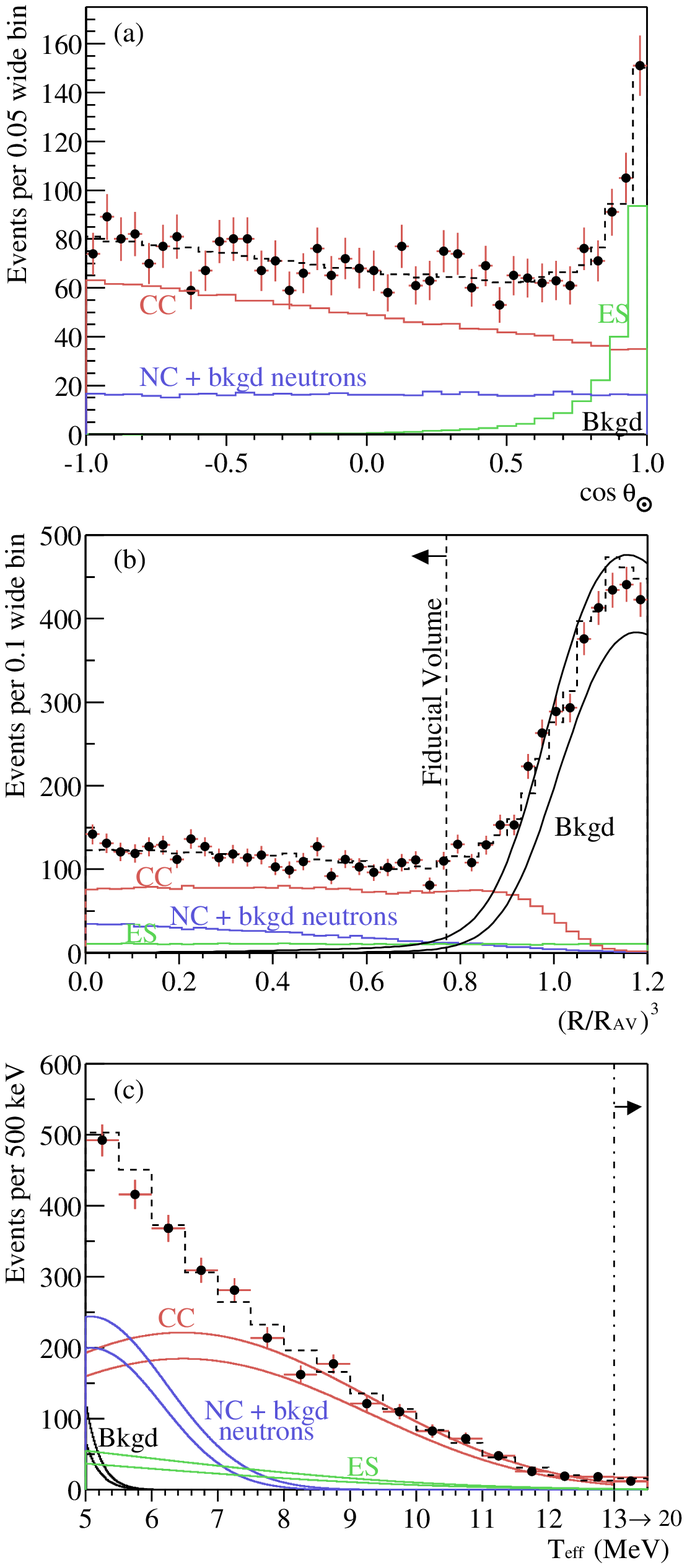}
\label{sno-0204008-f02a}
}
\hfill
\subfigure[
SNO electron kinetic energy spectrum.
]{
\includegraphics*[bb=6 5 324 244, width=0.47\textwidth]{fig/sno-0204008-f02.eps}
\label{sno-0204008-f02c}
}
\end{center}
\caption{ \label{sno-0204008-f02}
Figures taken from Ref.~\protect\cite{Ahmad:2002jz}.
}
\end{figure}
\begin{figure}[H]
\begin{center}
\includegraphics*[bb=2 10 523 517, height=0.40\textheight]{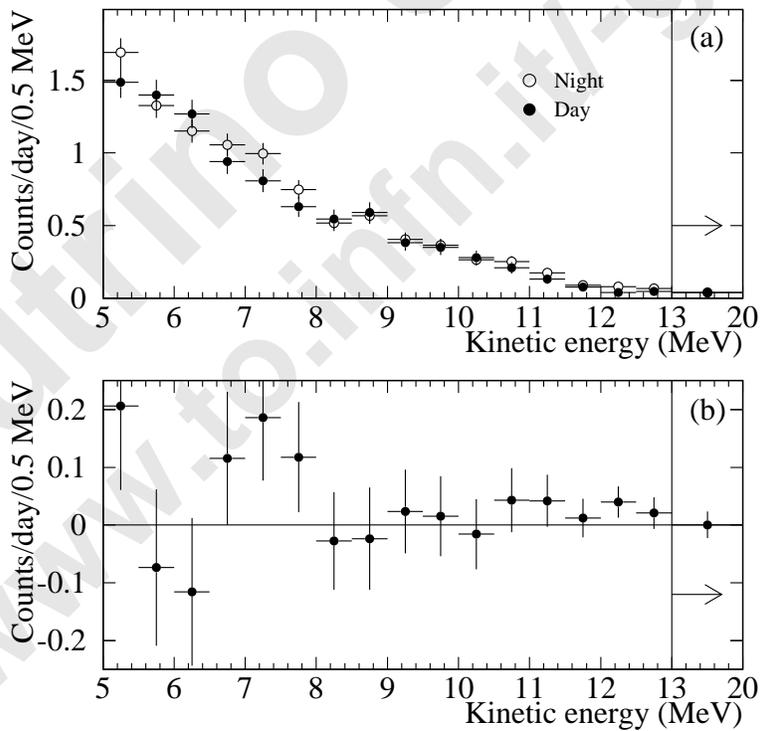}
\end{center}
\caption{ \label{sno-0204009-f02}
(a) SNO day and night energy spectra.
(b) Night $-$ day difference between the spectra
(night rate = $9.79 \pm 0.24$ events/day,
day rate = $9.23 \pm 0.27$ events/day).
Figure from Ref.~\protect\cite{Ahmad:2002ka}.
}
\end{figure}

\newpage

\section{Main characteristics of solar $\nu$ data}
\label{Main characteristics of solar nu data}
%
%
\footnotesize
\begin{center}
\violascuro
\setlength{\arraycolsep}{0pt}
\begin{tabular}{|c|c|c|c|c|c|}
\hline
Experiment
&
Reaction
&
$
\begin{array}{c}
E_{\mathrm{th}}
\\
\text{(MeV)}
\end{array}
$
&
$
\begin{array}{c}
\text{$\nu$ Flux}
\\
\text{Sensitivity}
\end{array}
$
&
$
\begin{array}{c}
\text{Operating}
\\
\text{Time}
\end{array}
$
&
$
\displaystyle
\frac
{ R^{\mathrm{exp}} }
{ R^{\mathrm{BP2000}} }
$
\\
\hline
\hline
SAGE
&
&
&
&
$1990-2001$
&
$ 0.55 \pm 0.05 $
\text{\verdescuro{\protect\cite{Abdurashitov:2002nt}}}
\vphantom{$\Big|$}
\\
\cline{1-1}
\cline{5-6}
GALLEX
&
$
\nu_e + {}^{71}\mathrm{Ga} \to {}^{71}\mathrm{Ge} + e^-
$
&
$ 0.233 $
&
\begin{tabular}{c}
\vphantom{$\Big|$}
$pp$,
${^7\mathrm{Be}}$,
${^8\mathrm{B}}$,
\\
\vphantom{$\Big|$}
$pep$,
$hep$,
\\
\vphantom{$\Big|$}
${^{13}\mathrm{N}}$,
${^{15}\mathrm{O}}$,
${^{17}\mathrm{F}}$
\end{tabular}
&
$1991-1997$
&
$ 0.61 \pm 0.06 $
\text{\verdescuro{\protect\cite{Hampel:1998xg}}}
\vphantom{$\Big|$}
\\
\cline{1-1}
\cline{5-6}
GNO
&
&
&
&
$1998-2000$
&
$ 0.51 \pm 0.08 $
\text{\verdescuro{\protect\cite{Altmann:2000ft}}}
\vphantom{$\Big|$}
\\
\hline
Homestake
&
$
\nu_e + ^{37}\mathrm{Cl}
\to
^{37}\mathrm{Ar} + e^-
$
&
$ 0.814 $
&
\begin{tabular}{c}
${^7\mathrm{Be}}$,
${^8\mathrm{B}}$,
\\
$pep$,
$hep$,
\\
${^{13}\mathrm{N}}$,
${^{15}\mathrm{O}}$,
${^{17}\mathrm{F}}$
\end{tabular}
&
$1970-1994$
&
$ 0.34 \pm 0.03 $
\text{\verdescuro{\protect\cite{Cleveland:1998nv}}}
\vphantom{$\Big|$}
\\
\hline
Kamiokande
&
&
$ 6.75 $
&
&
$
\begin{array}{c}
1987-1995
\\
\text{2079 days}
\end{array}
$
&
$ 0.55 \pm 0.08 $
\text{\verdescuro{\protect\cite{Fukuda:1996sz}}}
\vphantom{$\Big|$}
\\
\cline{1-1}
\cline{3-3}
\cline{5-6}
Super-Kam.
&
\raisebox{0.45cm}[0pt][0pt]
{$ \nu + e^- \to \nu + e^- $}
&
$ 4.75 $
&
&
$
\begin{array}{c}
1996-2001
\\
\text{1496 days}
\end{array}
$
&
$ 0.465 \pm 0.015 $
\text{\verdescuro{\protect\cite{Fukuda:2002pe}}}
\vphantom{$\Big|$}
\\
\cline{1-3}
\cline{5-6}
&
$ \nu_e + d \to p + p + e^- $
&
$ 6.9 $
&
${^8\mathrm{B}}$
&
&
$ 0.35 \pm 0.02 $
\text{\verdescuro{\protect\cite{Ahmad:2002jz}}}
\vphantom{$\Big|$}
\\
\cline{2-3}
\cline{6-6}
SNO
&
$ \nu + d \to p + n + \nu $
&
$ 2.2 $
&
&
$
\begin{array}{c}
1999-2002
\\
\text{306.4 days}
\end{array}
$
&
$ 1.01 \pm 0.13 $
\text{\verdescuro{\protect\cite{Ahmad:2002jz}}}
\vphantom{$\Big|$}
\\
\cline{2-3}
\cline{6-6}
&
$ \nu + e^- \to \nu + e^- $
&
$ 5.2 $
&
&
&
$ 0.47 \pm 0.05 $
\text{\verdescuro{\protect\cite{Ahmad:2002jz}}}
\vphantom{$\Big|$}
\\
\hline
\end{tabular}
\end{center}

\normalsize

\begin{equation}
\text{\violascuro{
Super-Kamiokande:
}}
\blue
\Phi_{hep} < 7.9 \, \Phi_{hep}^{\mathrm{SSM}} \quad \text{(90\% CL)}
\qquad
\text{\verdescuro{\protect\cite{Smy:2002rz}}}
\end{equation}
\begin{equation}
\text{\violascuro{
Super-Kamiokande energy spectrum:
no distorsion
\text{\verdescuro{\protect\cite{Smy:2002rz}}}
}}
\end{equation}
\begin{equation}
\text{\violascuro{
SNO energy spectrum:
no distorsion
\text{\verdescuro{\protect\cite{Ahmad:2002ka}}}
}}
\end{equation}
\begin{equation}
\text{\violascuro{
Super-Kamiokande time variations:
none
\text{\verdescuro{\protect\cite{Smy:2002rz}}}
}}
\end{equation}
\begin{equation}
\text{\violascuro{
Super-Kamiokande night-day asymmetry:
}}
\qquad
\blue
\mathcal{A^{\mathrm{SK}}_{\mathrm{ND}}}
=
0.021 \pm 0.024
\qquad
\text{\verdescuro{\protect\cite{Fukuda:2002pe}}}
\end{equation}
\begin{equation}
\text{\violascuro{
SNO night-day asymmetry:
}}
\qquad
\blue
\mathcal{A^{\mathrm{SNO}}_{\mathrm{ND}}}
=
0.070 \pm 0.051
\qquad
\text{\verdescuro{\protect\cite{Ahmad:2002ka}}}
\end{equation}

\newpage

\section{Solar neutrino transitions}
\label{Solar neutrino transitions}
%
%
\begin{description}
\item[Books:]
\protect\cite{Boehm:1992nn,Bahcall:1989ks,CWKim-book,Mohapatra:1998rq}
\item[Reviews:]
\protect\cite{Bilenky-Pontecorvo-PR-78,Bilenky:1987ty,Mikheev:1987qk,Kuo:1989qe,Pulido:1992fb,%
BGG-review-98,Fisher:1999fb,Bilenky:2001qi,Bilenkii:2001yh,Gonzalez-Garcia:2002dz}
\item[Vacuum Oscillations:]
\protect\cite{Pontecorvo:1968fh,Gribov:1969kq,Barger:1981xs,Acker:1991zj,Krastev:1992tj,Krastev:1993zx}
\item[MSW Effect:]
\protect\cite{Wolfenstein:1978ue,Barger:1980tf,Mikheev:1985gs,Mikheev:1986wj,%
Bethe:1986ej,Parke:1986jy,Petcov:1987xd,Petcov:1988zj,Krastev:1988ci,Petcov:1988wv,Kuo:1989pn}
\item[Regeneration in Earth:]
\protect\cite{Mikheev:1987qk,Cribier:1986ak,Baltz:1987hn,Baltz:1988sv,Baltz:1994fn,Lisi:1997yc,%
Liu:1997br,Petcov:1998su,Akhmedov:1998ui,Dighe:1999id,Chizhov:1999az,Chizhov:1999he,Guth:1999pi}
\item[Quasi-Vacuum Oscillations:]
\protect\cite{Friedland:2000cp,Fogli:2000bk,Friedland:2000rn,Lisi:2000su}
\item[Three-Neutrino Mixing:]
\protect\cite{Kuo:1986sk,Kuo:1987zx,Toshev:1987fs,Petcov:1987qg,Shi:1992zw}
\item[Four-Neutrino Mixing:]
\protect\cite{Dooling:1999sg,Giunti:2000wt}
\item[Flavor-Changing Neutral Currents:]
\protect\cite{Wolfenstein:1978ue,Valle:1987gv,Guzzo:1991hi,Guzzo:1991cp}
\item[Spin-Flavor Precession:]
\protect\cite{Cisneros:1971nq,Voloshin:1986ty,Okun:1986uf,Okun:1986na,Okun:1986hi,Akhmedov:1988nc,Lim:1988tk,Akhmedov:1988uk,Akhmedov:1989df}
\end{description}

\newpage

\section{Two-neutrino oscillations in vacuum and matter}
\label{Two-neutrino oscillations in vacuum and matter}

\begin{equation}
\text{mixing:}
\quad
\nu_e
=
\cos\!\vartheta \nu_1
+
\sin\!\vartheta \nu_2
\,,
\quad
\nu_f
=
-
\sin\!\vartheta \nu_1
+
\cos\!\vartheta \nu_2
\quad
(f = \mu, \tau, s)
\label{M00}
\end{equation}
\begin{equation}
\text{transition probability in vacuum:}
\quad
P_{\nu_e\to\nu_f}(R)
=
\sin^2\!2\vartheta
\sin^2\left( \frac{\Delta{m}^2 R}{4E} \right)
\label{M00a}
\end{equation}
\begin{equation}
R = \text{distance from the center of the Sun}
\,,
\quad
\Delta{m}^2 \equiv m_2^2 - m_1^2
\label{M00b}
\end{equation}
\begin{equation}
\begin{array}{l}
\text{evolution}
\\
\text{in matter:}
\end{array}
\quad
i \frac{ \mathrm{d} }{ \mathrm{d}R }
\begin{pmatrix}
\phi_{\nu_e}(R)
\\
\phi_{\nu_f}(R)
\end{pmatrix}
=
\frac{1}{4E}
\begin{pmatrix}
- \Delta{m}^2 \cos\!2\vartheta + 2 A
&
\Delta{m}^2 \sin\!2\vartheta
\\
\Delta{m}^2 \sin\!2\vartheta
&
\Delta{m}^2 \cos\!2\vartheta
\end{pmatrix}
\begin{pmatrix}
\phi_{\nu_e}(R)
\\
\phi_{\nu_f}(R)
\end{pmatrix}
\label{M01}
\end{equation}
\begin{equation}
\phi_{\nu_e}(0) = 1
\,,
\quad
\phi_{\nu_f}(0) = 0
\quad
\Longrightarrow
\quad
P_{\nu_e\to\nu_f}(R)
=
|\phi_{\nu_f}(R)|^2
\label{M01a}
\end{equation}
\begin{equation}
A = 2 E V
\quad
\text{with}
\quad
\left\{
\begin{array}{ll}
V = V_{CC} = \sqrt{2} G_F N_e
&
\quad \text{for} \quad
f = \mu, \tau
\\
V = V_{CC} + V_{NC} = \sqrt{2} G_F \left( N_e - \frac{1}{2} N_n \right)
&
\quad \text{for} \quad
f = s
\end{array}
\right.
\label{M02}
\end{equation}
\begin{center}
$ N_e = \text{electron number density}$
\hspace{2cm}
$ N_n = \text{neutron number density}$
\end{center}
\begin{figure}[H]
\begin{center}
\subfigure[
$
V_{CC}
=
\sqrt{2} G_{F} N_{e}
$.
]{
\includegraphics*[bb=189 540 422 770, width=0.4\textwidth]{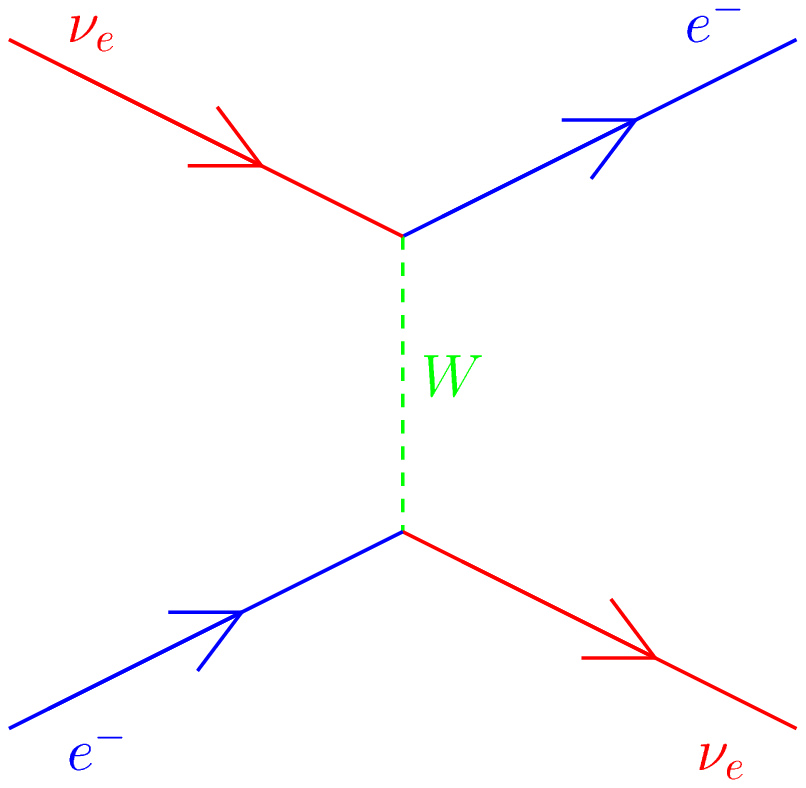}
\label{vcc}
}
\hfill
\subfigure[
$
V_{NC}^{(e^-)}
=
-
V_{NC}^{(p)}
$;
$
V_{NC}
=
V_{NC}^{(n)}
=
-
\frac{ \sqrt{2} }{ 2 } 
G_{F} N_{n}
$.
]{
\includegraphics*[bb=189 540 422 770, width=0.4\textwidth]{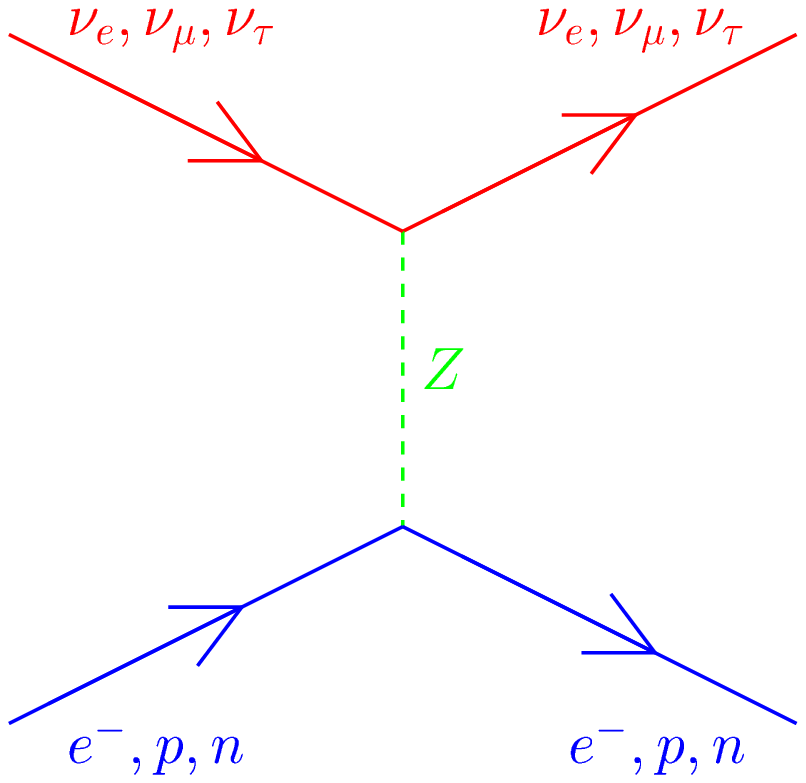}
\label{vnc}
}
\end{center}
\caption{ \label{veff}
$
V_{\nu_e}
=
V_{CC} + V_{NC}
$;
$
V_{\nu_\mu}
=
V_{\nu_\tau}
=
V_{NC}
$;
$
V_{\nu_e} - V_{\nu_{\mu,\tau}}
=
V_{CC}
=
\sqrt{2} G_{F} N_{e}
$;
$
V_{\nu_e} - V_{\nu_s}
=
V_{CC} + V_{NC}
=
\sqrt{2} G_{F} \left( N_{e} - \frac{1}{2} N_{n} \right)
$.
}
\end{figure}
\begin{equation}
V
=
\sqrt{2} G_F N
=
7.63 \times 10^{-14}
\left( \frac{N}{N_A \, \mathrm{cm}^{-3}} \right) \, \mathrm{eV}
\quad
\text{with}
\quad
\left\{
\begin{array}{ll}
N = N_e
&
\quad \text{for} \quad
f = \mu, \tau
\\
N = N_e - \frac{1}{2} N_n
&
\quad \text{for} \quad
f = s
\end{array}
\right.
\label{M04c}
\end{equation}
\begin{equation}
\text{effective mixing angle:}
\quad
\tan\!2\vartheta_M
=
\frac{\tan\!2\vartheta}{1-\frac{A}{\Delta{m}^2\cos\!2\vartheta}}
\,,
\quad
\text{resonance:}
\quad
A_{\mathrm{res}}=\Delta{m}^2\cos\!2\vartheta
\label{M03}
\end{equation}
\begin{equation}
\text{effective squared-mass difference:}
\quad
\Delta{m}^2_M
=
\sqrt{
\left( \Delta{m}^2\cos\!2\vartheta - A \right)^2
+
\left( \Delta{m}^2\sin\!2\vartheta \right)^2
}
\label{M04}
\end{equation}
\begin{equation}
\text{effective squared masses:}
\quad
(m^M_{2,1})^2
=
m_1^2 + \frac{1}{2} \left( \Delta{m}^2 + A \pm \Delta{m}^2_M \right)
\label{M04b}
\end{equation}

\newpage

standard terminology for regions in the $\Delta{m}^2$--$\tan^2\vartheta$ plane:
\begin{align}
\text{LMA (Large Mixing Angle):}
\null & \null
\quad
\Delta{m}^2 \sim 5 \times 10^{-5} \, \mathrm{eV}^2
\,,
\null && \null
\tan^2\vartheta \sim 0.8
\label{LMA}
\\
\text{LOW (LOW $\Delta{m}^2$):}
\null & \null
\quad
\Delta{m}^2 \sim 7 \times 10^{-8} \, \mathrm{eV}^2
\,,
\null && \null
\tan^2\vartheta \sim 0.6
\label{LOW}
\\
\text{SMA (Small Mixing Angle):}
\null & \null
\quad
\Delta{m}^2 \sim 5 \times 10^{-6} \, \mathrm{eV}^2
\,,
\null && \null
\tan^2\vartheta \sim 10^{-3}
\label{SMA}
\\
\text{QVO (Quasi-Vacuum Oscillations):}
\null & \null
\quad
\Delta{m}^2 \sim 10^{-9} \, \mathrm{eV}^2
\,,
\null && \null
\tan^2\vartheta \sim 1
\label{QVO}
\\
\text{VAC (VACuum oscillations):}
\null & \null
\quad
\Delta{m}^2 \lesssim 5 \times 10^{-10} \, \mathrm{eV}^2
\,,
\null && \null
\tan^2\vartheta \sim 1
\label{VAC}
\end{align}
\begin{figure}[H]
\begin{center}
\subfigure[
Figure from Ref.~\protect\cite{deGouvea:2000cq}.
]{
\includegraphics*[bb=0 0 435 572, width=0.40\textwidth]{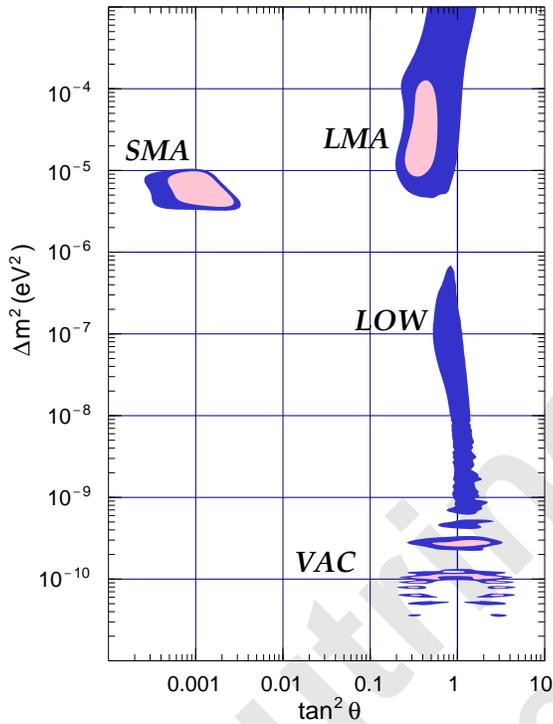}
\label{degouvea-0002064-f01}
}
\hfill
\subfigure[
Figure from Ref.~\protect\cite{Bahcall:2001hv}.
]{
\includegraphics*[bb=70 205 523 644, width=0.55\textwidth]{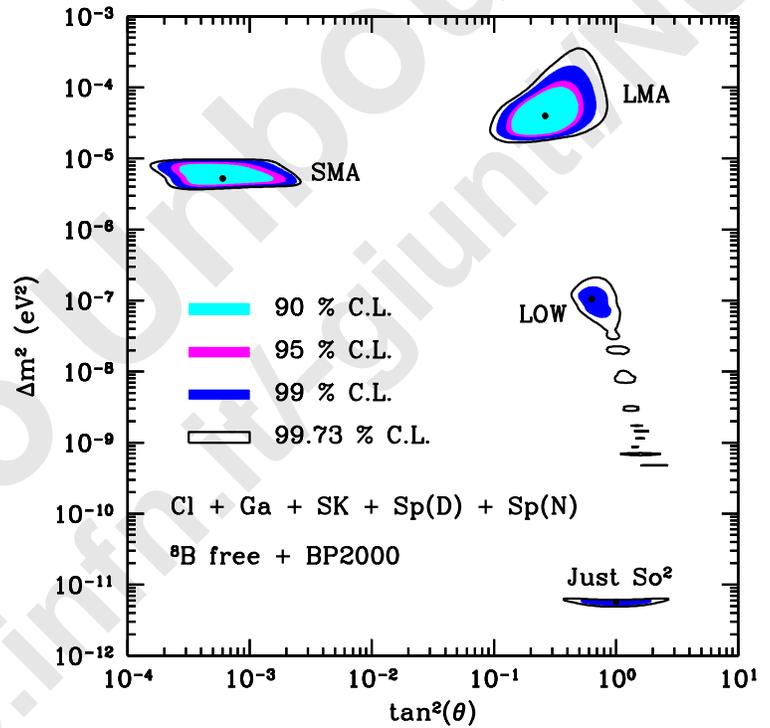}
\label{bahcall-0103179-f01a}
}
\end{center}
\caption{ \label{regions}
Regions in the $\Delta{m}^2$--$\tan^2\vartheta$ plane.
}
\end{figure}

\newpage

\begin{figure}[H]
\begin{center}
\includegraphics*[bb=66 429 400 753, width=0.45\textwidth]{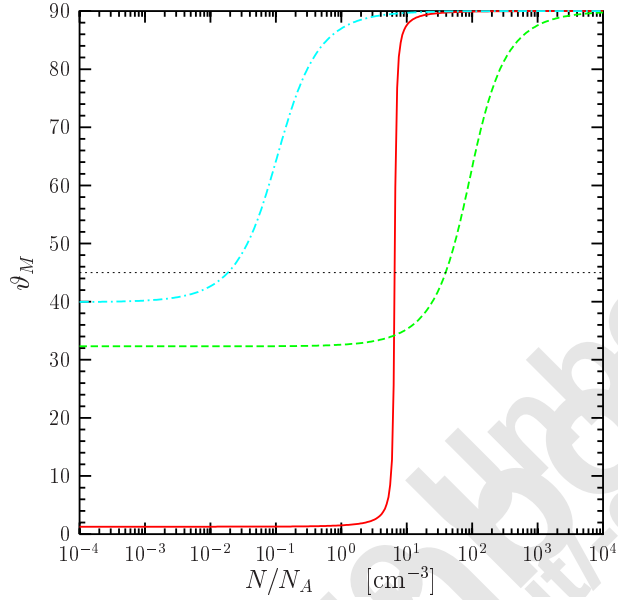}
\end{center}
\caption{ \label{tm}
Effective mixing angle in matter
as a function of
the effective number density $N=N_e$
for $\nu_e\to\nu_{\mu,\tau}$ transitions
and
$N=N_e-N_n/2$
for $\nu_e\to\nu_s$ transitions.
Energy: $E = 5 \, \mathrm{MeV}$.
Solid line:
$
\Delta{m}^2 = 5 \times 10^{-6} \, \mathrm{eV}^2
\,,
\,
\tan^2\vartheta = 5 \times 10^{-4}
$
(typical SMA).
Dashed line:
$
\Delta{m}^2 = 7 \times 10^{-5} \, \mathrm{eV}^2
\,,
\,
\tan^2\vartheta = 0.4
$
(typical LMA).
Dash-dotted line:
$
\Delta{m}^2 = 8 \times 10^{-8} \, \mathrm{eV}^2
\,,
\,
\tan^2\vartheta = 0.7
$
(typical LOW).
}
\end{figure}
\begin{figure}[H]
\begin{center}
\subfigure[
Typical SMA:
$
\Delta{m}^2 = 5 \times 10^{-6} \, \mathrm{eV}^2
\,,
\,
\tan^2\vartheta = 5 \times 10^{-4}
$.
]{
\includegraphics*[bb=76 428 410 750, width=0.31\textwidth]{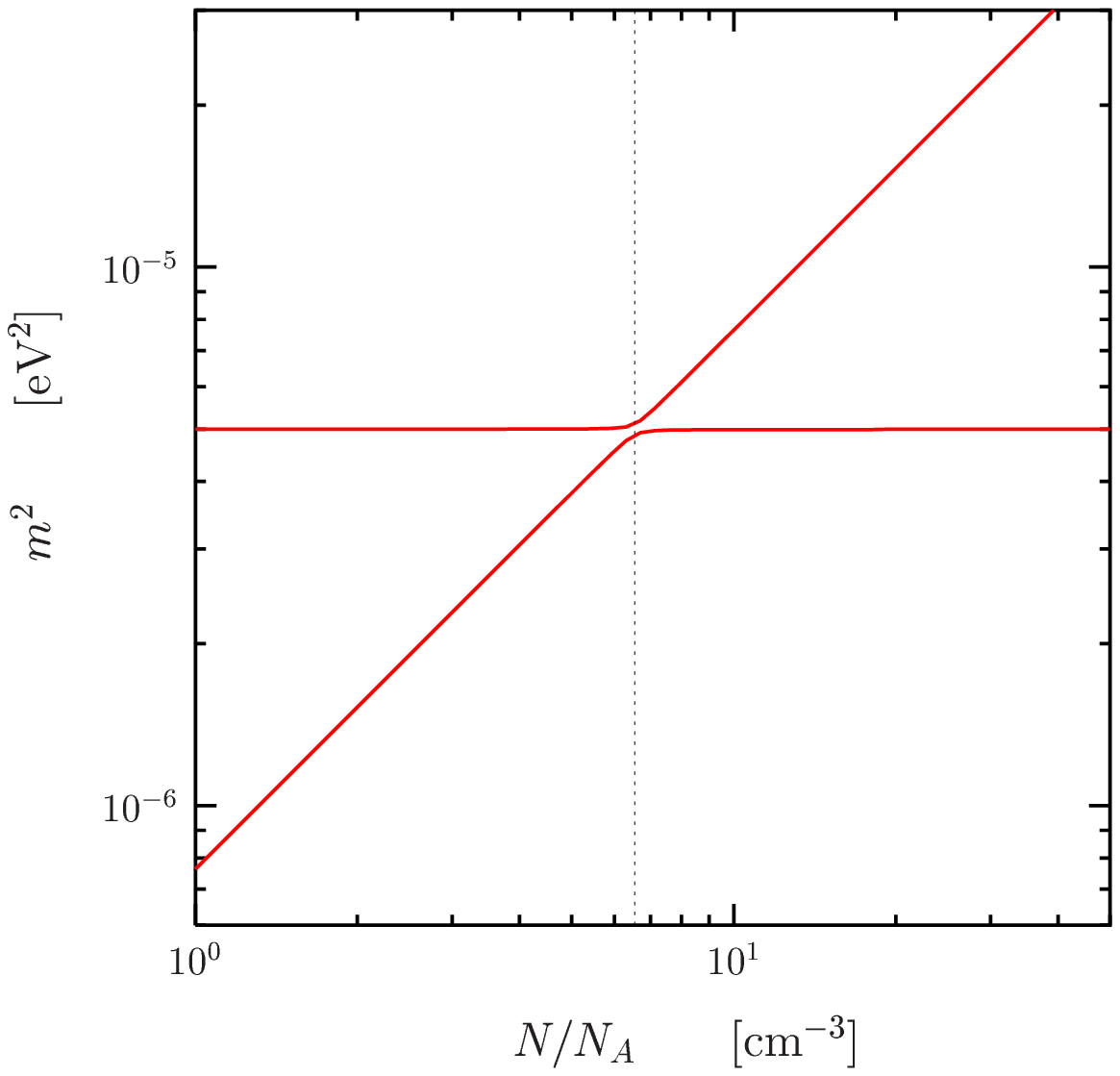}
\label{mm-sma}
}
\hfill
\subfigure[
Typical LMA:
$
\Delta{m}^2 = 7 \times 10^{-5} \, \mathrm{eV}^2
\,,
\,
\tan^2\vartheta = 0.4
$.
]{
\includegraphics*[bb=76 428 410 750, width=0.31\textwidth]{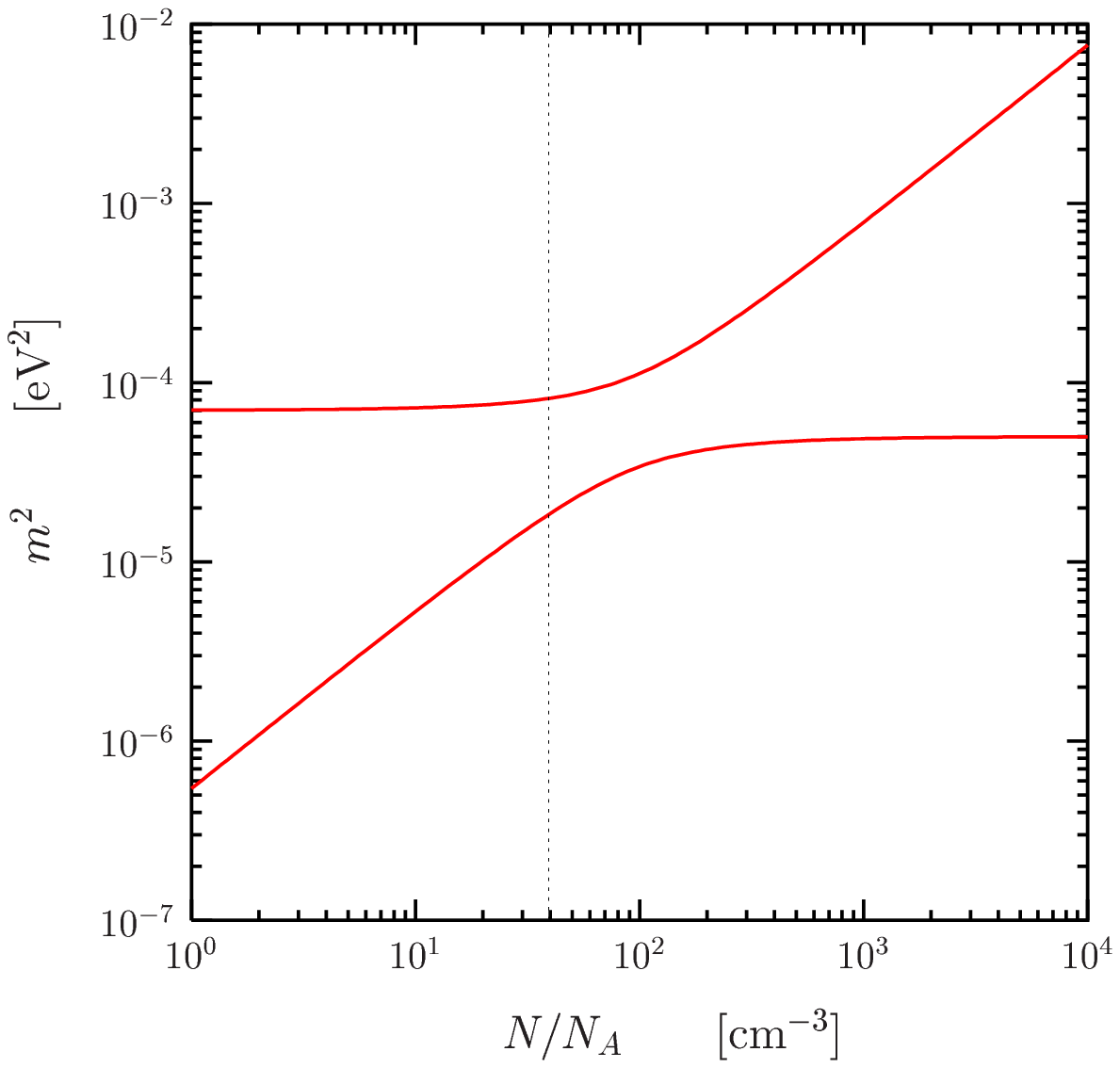}
\label{mm-lma}
}
\hfill
\subfigure[
Typical LOW:
$
\Delta{m}^2 = 8 \times 10^{-8} \, \mathrm{eV}^2
\,,
\,
\tan^2\vartheta = 0.7
$.
]{
\includegraphics*[bb=76 428 410 750, width=0.31\textwidth]{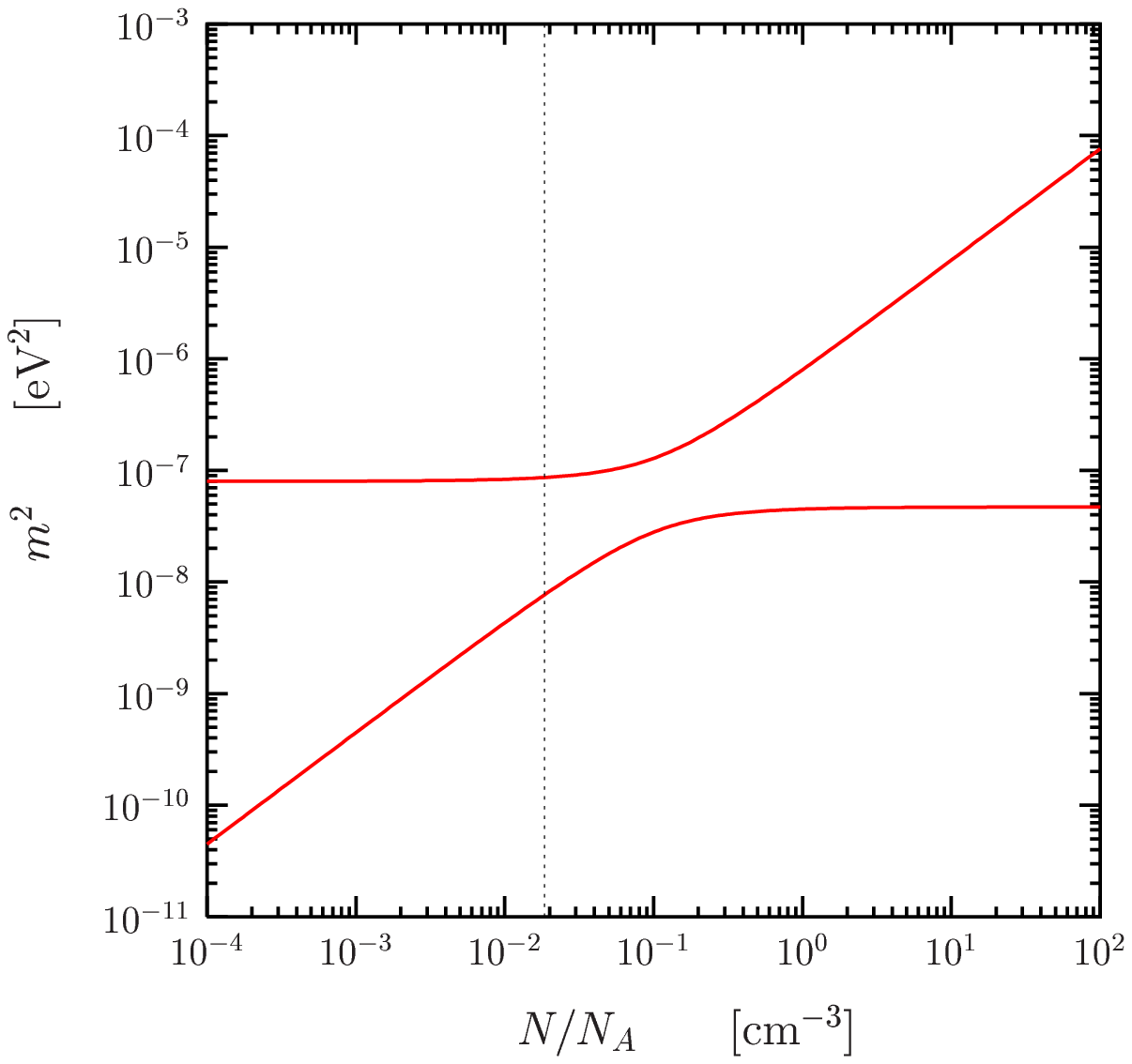}
\label{mm-low}
}
\end{center}
\caption{ \label{mm}
Effective squared masses in matter (\ref{M04b})
as functions of
the effective number density $N=N_e$
for $\nu_e\to\nu_{\mu,\tau}$ transitions
and
$N=N_e-N_n/2$
for $\nu_e\to\nu_s$ transitions, assuming $m_1=0$.
Energy: $E = 5 \, \mathrm{MeV}$.
The dotted vertical lines show the location of the resonance (Eq.~(\ref{M03})),
where the effective squared-mass difference $\Delta{m}^2_M$ in Eq.~(\ref{M04})
is minimal
(in Figs.~\ref{mm-lma} and \ref{mm-low}
the location of the resonance appears off-center because of the logaritmic scale).
}
\end{figure}

\newpage

exponential approximation of electron number density in the Sun \protect\cite{Bahcall:1989ks}:
\begin{equation}
N_e(R)
=
N_e(0) \exp\left( - \frac{R}{R_0} \right)
\,,
\quad
N_e(0)
=
245 \, \mathrm{mol} / \mathrm{cm}^{3}
\,,
\quad
R_0 = \frac{R_\odot}{10.54}
\label{M05}
\end{equation}
\begin{figure}[H]
\begin{center}
\subfigure[
The electron number density, $n_e=N_e$, versus solar radius in the BP2000 SSM.
The straight line represents the exponential approximation in Eq.~(\ref{M05}).
]{
\rotatebox{-90}{
\includegraphics*[bb=40 25 582 742, height=0.45\textwidth]{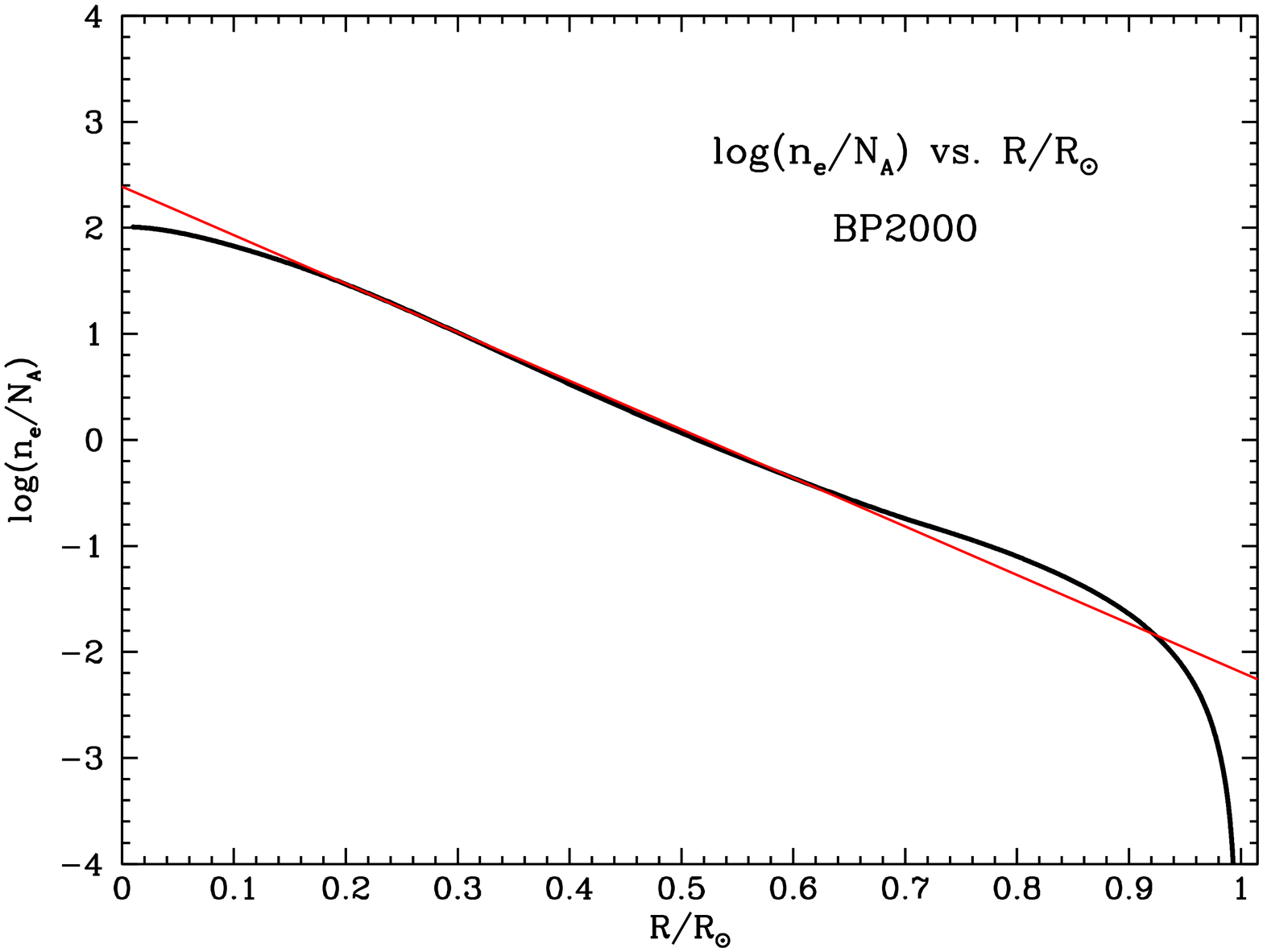}
}
\label{bahcall-0010346-f08}
}
\hfill
\subfigure[
Number density of scatterers
$n_\mathrm{sterile}=N_e-N_n/2$
relevant for
$\nu_e \to \nu_s$ transitions
versus solar radius in the BP2000 SSM.
]{
\rotatebox{-90}{
\includegraphics*[bb=40 25 582 742, height=0.45\textwidth]{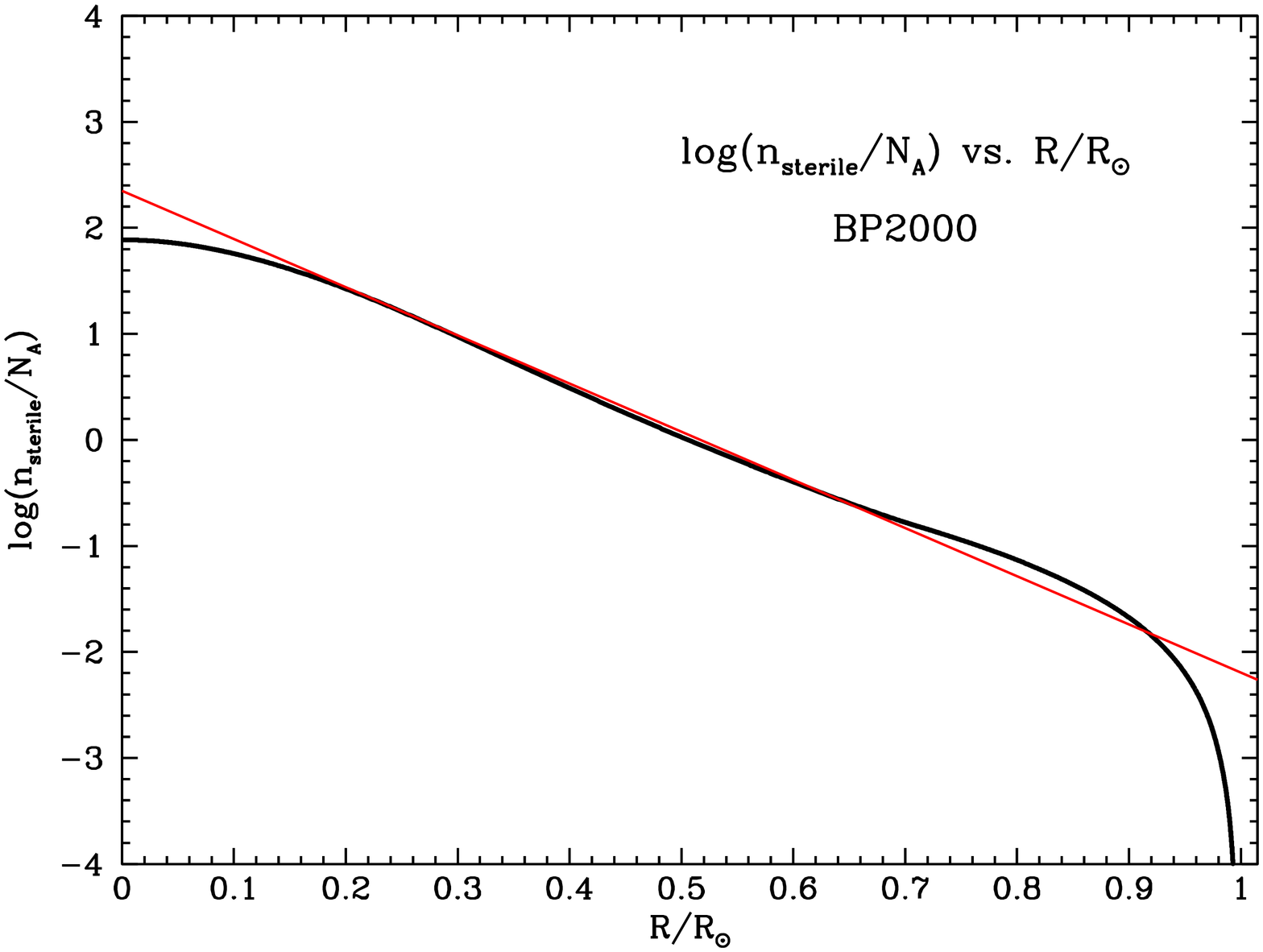}
}
\label{bahcall-0010346-f09}
}
\end{center}
\caption{ \label{neff}
Figures taken from Ref.~\protect\cite{Bahcall:2000nu}.
Precise numerical values are available at Ref.~\protect\cite{Bahcall-WWW}.
}
\end{figure}
average $\nu_e$ survival probability after MSW transitions in Sun (Parke formula) \protect\cite{Parke:1986jy}:
\begin{equation}
P_{\nu_e\to\nu_e}^{\mathrm{sun}}
=
\frac{1}{2}
+
\left( \frac{1}{2} - P_c \right) \cos\!2\vartheta \cos\!2\vartheta_{M}^{0}
\,,
\qquad
\vartheta_{M}^{0} = \text{effective mixing angle at production}
\label{M06}
\end{equation}
\begin{equation}
\text{$\nu_1\leftrightarrows\nu_2$ crossing probability
\protect\cite{Kuo:1989pn}:}
\qquad
P_c
=
\frac
{
\exp\left( - \frac{\pi}{2} \gamma F \right)
-
\exp\left( - \frac{\pi}{2} \gamma \frac{F}{\sin^2\vartheta} \right)
}
{
1
-
\exp\left( - \frac{\pi}{2} \gamma \frac{F}{\sin^2\vartheta} \right)
}
\label{M07}
\end{equation}
\begin{equation}
\gamma
=
\frac
{\Delta{m}^2 \sin^22\vartheta}
{2 E \cos\!2\vartheta \left|\frac{\mathrm{d}\ln{A}}{\mathrm{d}R}\right|_{\mathrm{res}}}
\label{M08}
\end{equation}
\begin{align}
\null & \null
\text{$A \propto R$ \protect\cite{Parke:1986jy,Haxton:1987bc,Pizzochero:1987fj,Petcov:1987xd,Kuo:1989pn,Balantekin:1998jp}:}
\null & \quad & \null
F = 1
\label{M09a}
\\
\null & \null
\text{$A \propto 1/R$ \protect\cite{Kuo:1989pn}:}
\null & \quad & \null
F = \left(1-\tan^2\vartheta\right)^2/\left(1+\tan^2\vartheta\right)
\label{M09b}
\\
\null & \null
\text{$A \propto \exp\left(-R\right)$ \protect\cite{Pizzochero:1987fj,Toshev:1987jw,Petcov:1988zj,Petcov:1988wv,Balantekin:1998jp}:}
\null & \quad & \null
F = 1 - \tan^2\vartheta
\label{M09c}
\end{align}
\begin{equation}
\text{practical prescription \protect\cite{Lisi:2000su}:
use Eq.~(\ref{M09c}) and }
\left\{
\begin{array}{ll}
\text{
numerical
$
\displaystyle
\left|\frac{\mathrm{d}\ln{A}}{\mathrm{d}R}\right|_{\mathrm{res}}
$
}
&
\text{
for
$R \leq 0.904 R_{\odot}$
}
\\
\text{
$
\displaystyle
\left|\frac{\mathrm{d}\ln{A}}{\mathrm{d}R}\right|_{\mathrm{res}} \to \frac{18.9}{R_{\odot}}
$
}
&
\text{
for
$R > 0.904 R_{\odot}$
}
\end{array}
\right.
\label{M10}
\end{equation}

\begin{equation}
\text{$\nu_e$ regeneration in Earth \protect\cite{Mikheev:1987qk,Baltz:1987hn}:}
\quad
P_{\nu_e\to\nu_e}^{\mathrm{sun+earth}}
=
P_{\nu_e\to\nu_e}^{\mathrm{sun}}
+
\frac
{
\left(1-2P_{\nu_e\to\nu_e}^{\mathrm{sun}}\right)
\left(P_{\nu_2\to\nu_e}^{\mathrm{earth}}-\sin^2\vartheta\right)
}
{\cos\!2\vartheta}
\label{M11}
\end{equation}

\newpage

\begin{figure}[H]
\begin{center}
\subfigure[
SMA:
$
\Delta{m}^2 = 5.0 \times 10^{-6} \, \mathrm{eV}^2
\,,
\,
\sin^2\!2\vartheta = 3.5 \times 10^{-3}
$.
LMA:
$
\Delta{m}^2 = 1.6 \times 10^{-5} \, \mathrm{eV}^2
\,,
\,
\sin^2\!2\vartheta = 0.57
$.
LOW:
$
\Delta{m}^2 = 7.9 \times 10^{-8} \, \mathrm{eV}^2
\,,
\,
\sin^2\!2\vartheta = 0.95
$.
Figure from Ref.~\protect\cite{Bahcall:1998jt}.
]{
\includegraphics*[bb=133 188 467 679, width=0.42\textwidth]{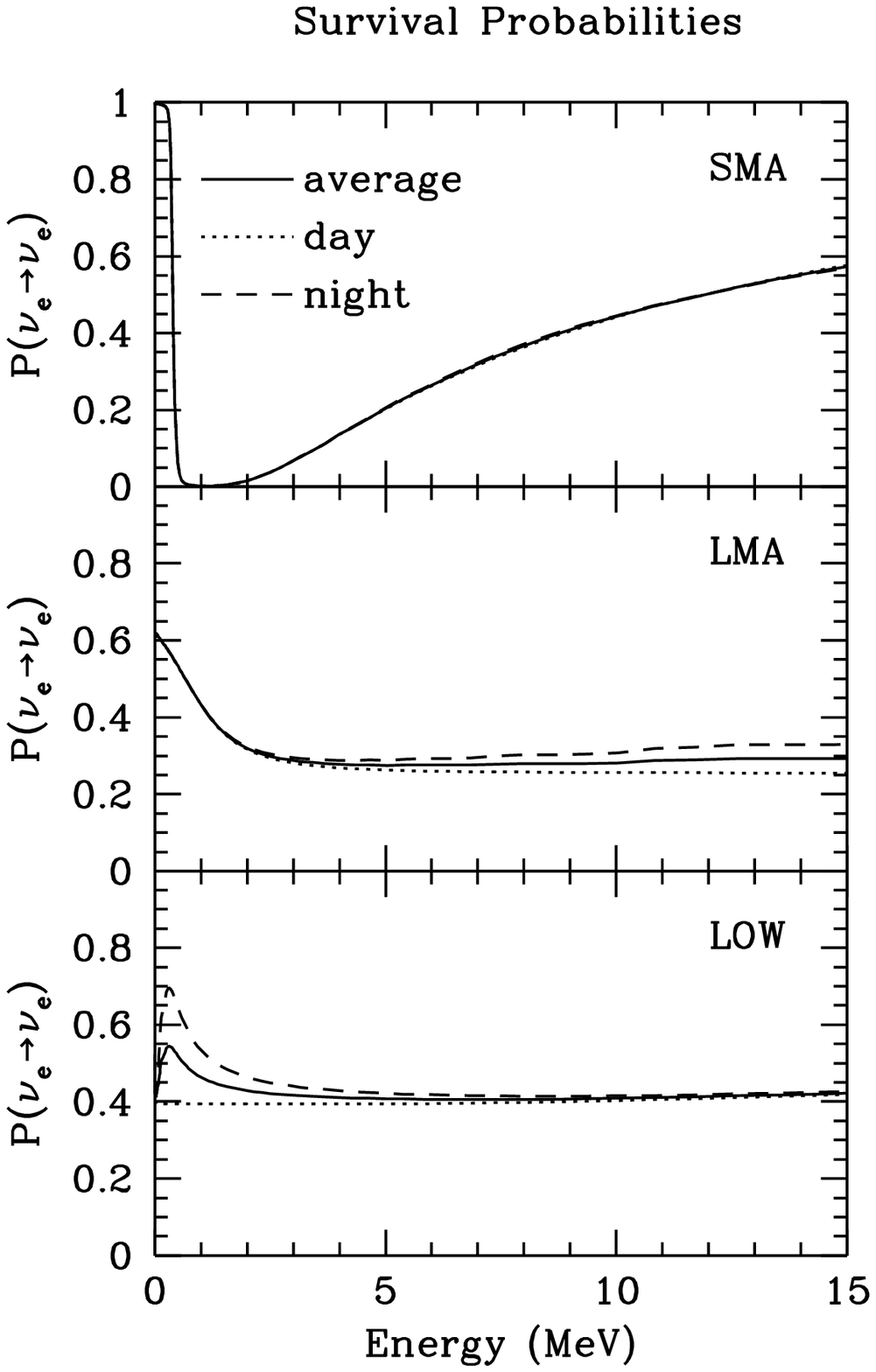}
\label{bahcall-9807216-f09}
}
\hfill
\subfigure[
LMA:
$
\Delta{m}^2 = 4.2 \times 10^{-5} \, \mathrm{eV}^2
\,,
\,
\tan^2\vartheta = 0.26
$.
SMA:
$
\Delta{m}^2 = 5.2 \times 10^{-6} \, \mathrm{eV}^2
\,,
\,
\tan^2\vartheta = 5.5 \times 10^{-4}
$.
LOW:
$
\Delta{m}^2 = 7.6 \times 10^{-8} \, \mathrm{eV}^2
\,,
\,
\tan^2\vartheta = 0.72
$.
Just So$^2$:
$
\Delta{m}^2 = 5.5 \times 10^{-12} \, \mathrm{eV}^2
\,,
\,
\tan^2\vartheta = 1.0
$.
VAC:
$
\Delta{m}^2 = 1.4 \times 10^{-10} \, \mathrm{eV}^2
\,,
\,
\tan^2\vartheta = 0.38
$.
Figure from Ref.~\protect\cite{Bahcall:2001hv}.
]{
\includegraphics*[bb=25 82 576 714, width=0.535\textwidth]{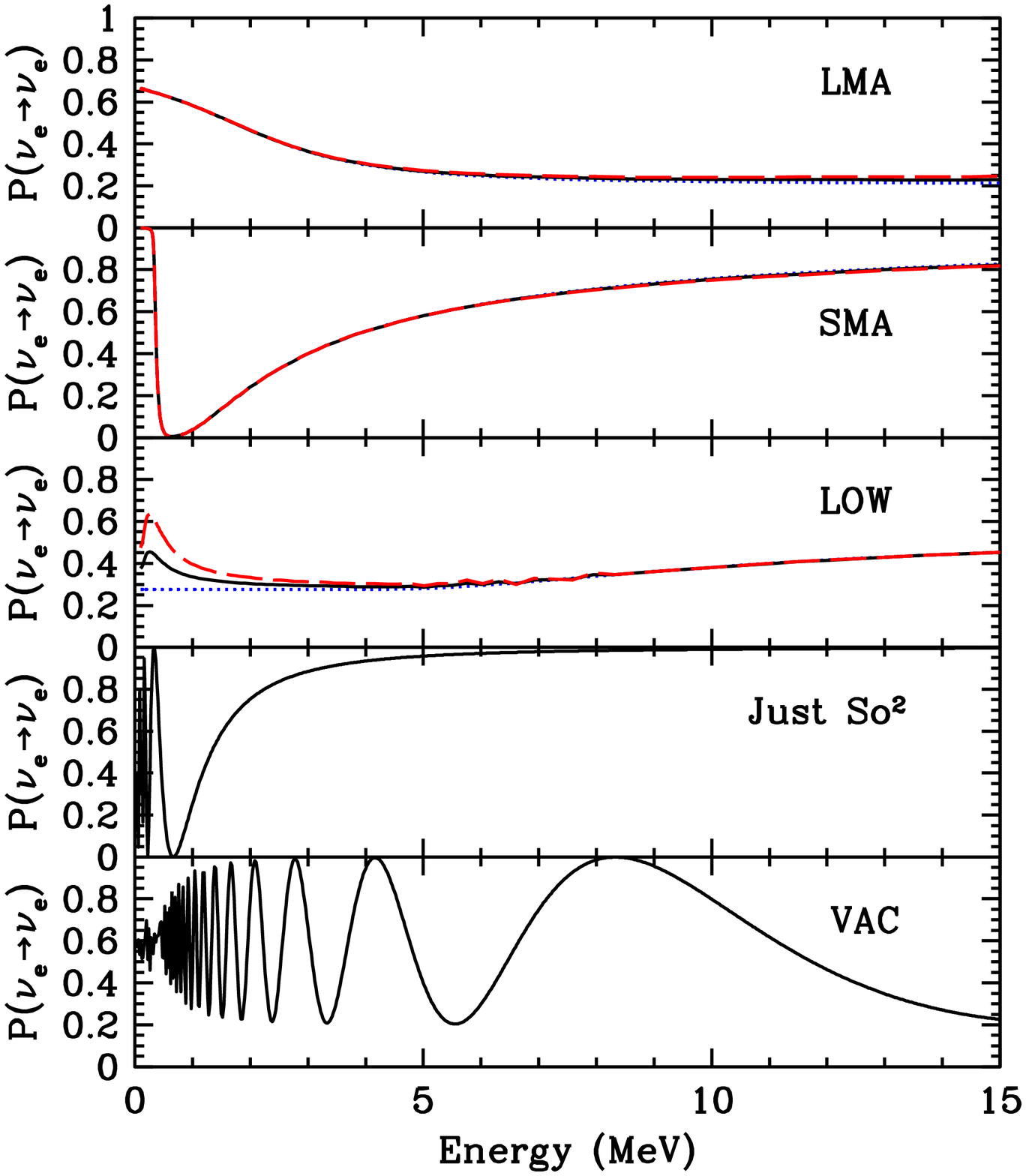}
\label{bahcall-0103179-f02}
}
\end{center}
\caption{ \label{Pee}
Solar $\nu_e$ survival probability as a function of energy.
Regeneration in the Earth is included.
}
\end{figure}
\begin{figure}[H]
\begin{center}
\subfigure[
LMA:
$
\Delta{m}^2 = 3.7 \times 10^{-5} \, \mathrm{eV}^2
\,,
\,
\sin^22\vartheta = 0.79
$.
LOW:
$
\Delta{m}^2 = 1.0 \times 10^{-7} \, \mathrm{eV}^2
\,,
\,
\sin^22\vartheta = 0.97
$.
GP:
Gribov-Pontecorvo solution
$P_{ee}=1/2$ \protect\cite{Gribov:1969kq}.
]{
\includegraphics*[bb=116 268 502 652, width=0.31\textwidth]{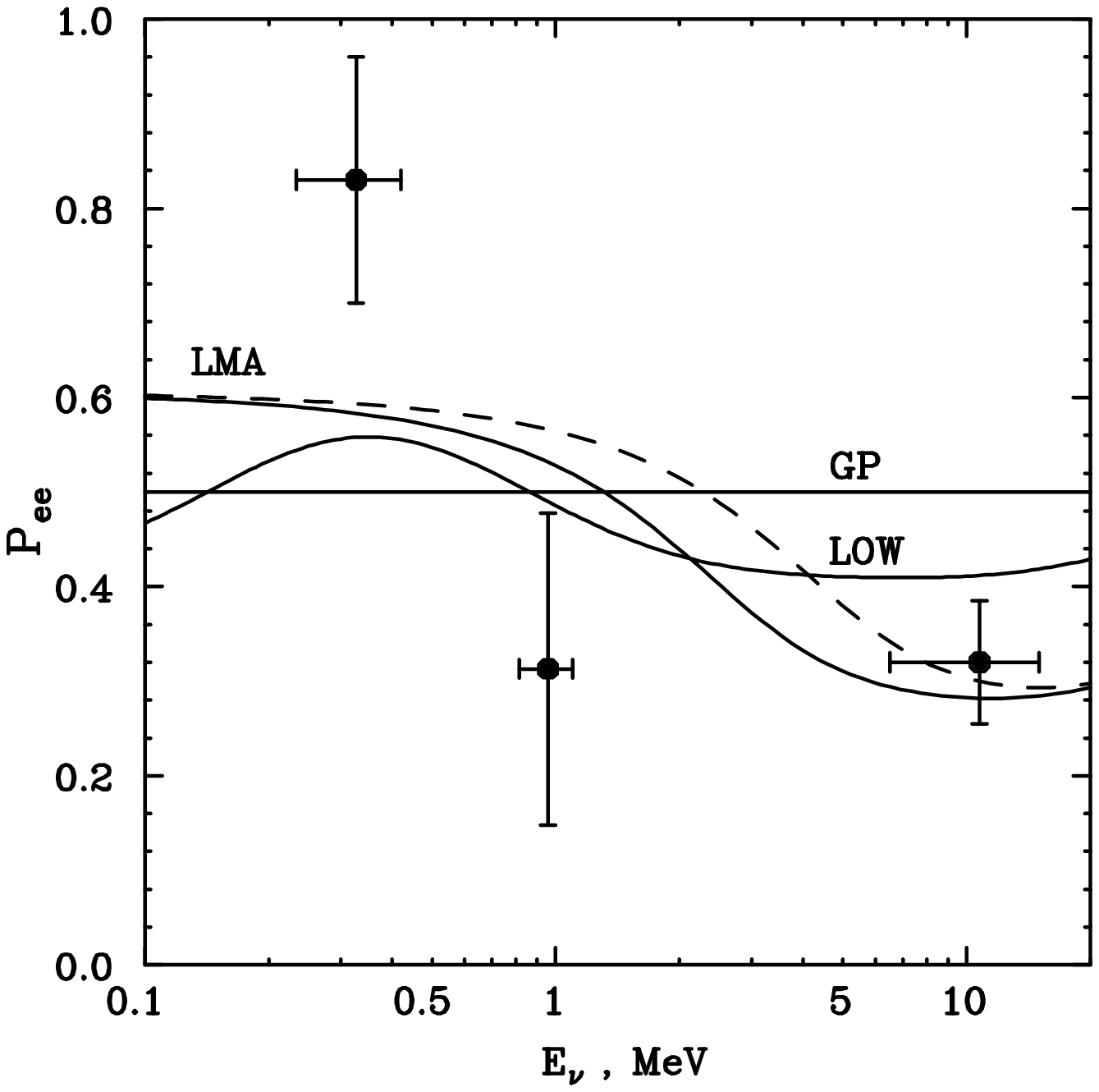}
\label{berezinsky-0108108-f01}
}
\hfill
\subfigure[
SMA:
$
\Delta{m}^2 = 4.6 \times 10^{-6} \, \mathrm{eV}^2
\,,
\,
\sin^22\vartheta = 1.4 \times 10^{-3}
$.
]{
\includegraphics*[bb=116 268 502 652, width=0.31\textwidth]{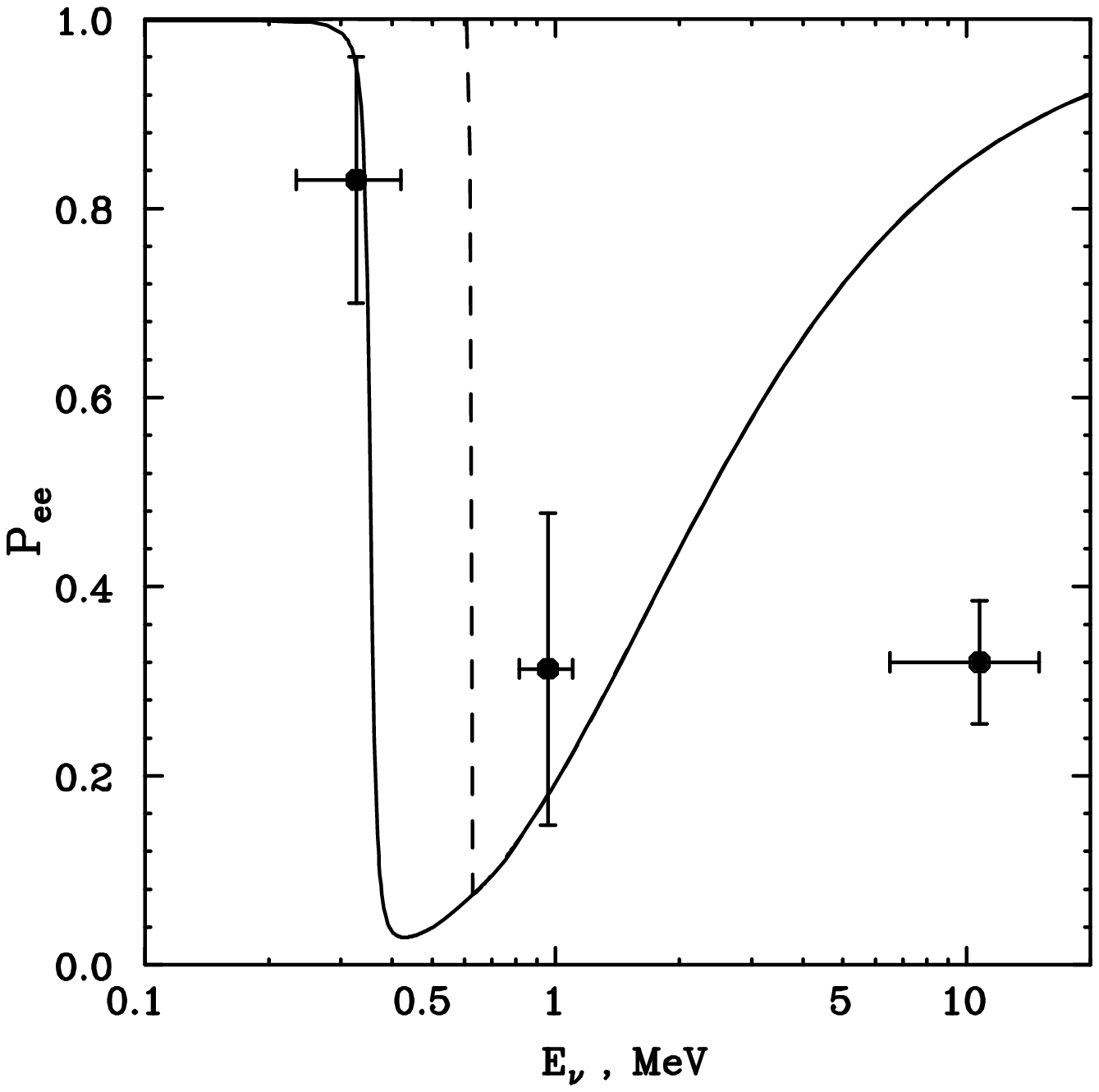}
\label{berezinsky-0108108-f02}
}
\hfill
\subfigure[
VAC (solid):
$
\Delta{m}^2 = 4.6 \times 10^{-10} \, \mathrm{eV}^2
\,,
\,
\sin^22\vartheta = 0.83
$.
Just So$^2$ (dashed):
$
\Delta{m}^2 = 5.5 \times 10^{-12} \, \mathrm{eV}^2
\,,
\,
\tan^2\vartheta = 0.96
$.
]{
\includegraphics*[bb=116 268 502 652, width=0.31\textwidth]{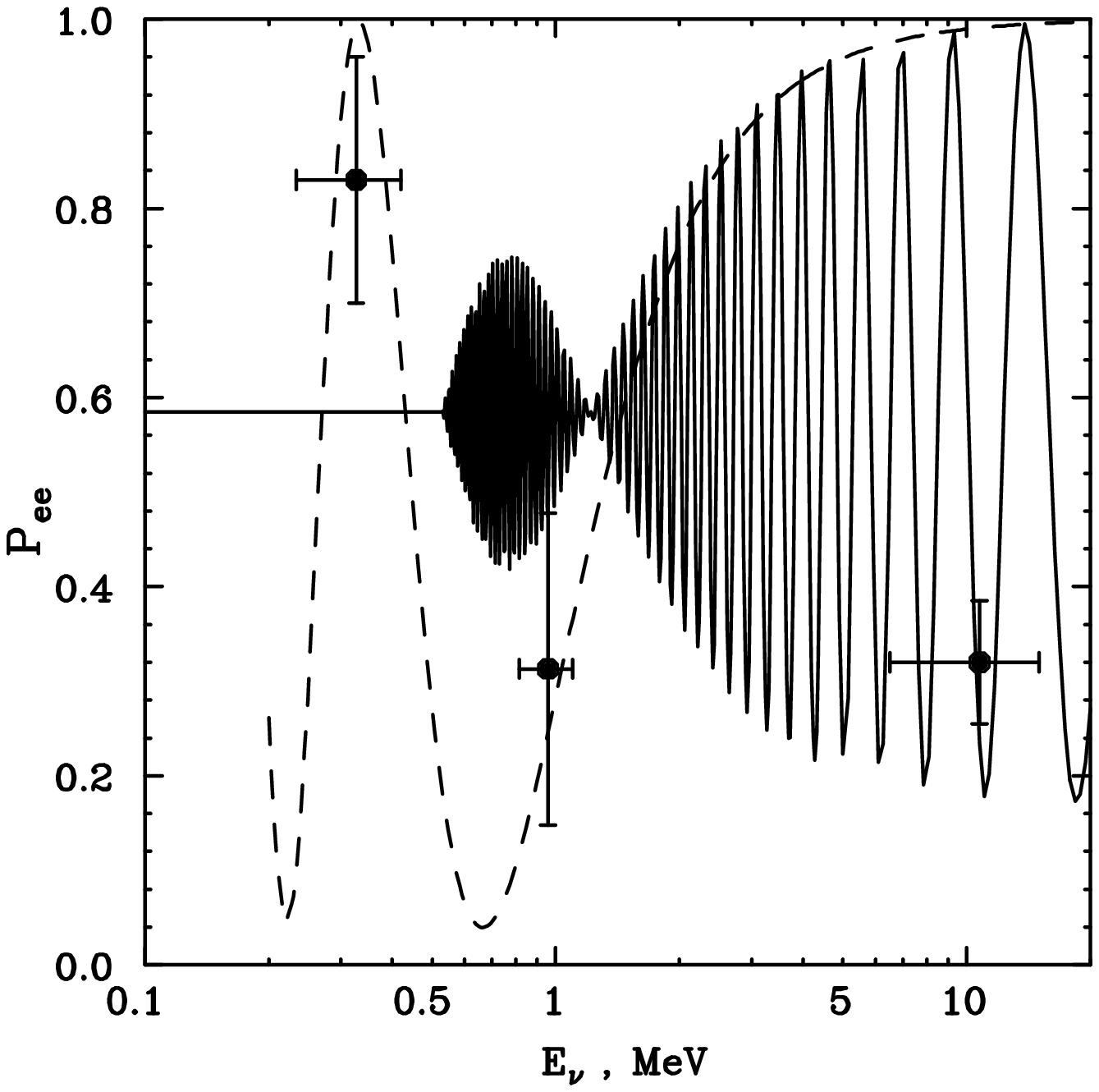}
\label{berezinsky-0108108-f03}
}
\end{center}
\caption{ \label{berezinsky-0108108}
Survival probability of electron neutrinos as a function
of energy. Data points are extracted from
the gallium, chlorine and boron-neutrino signals.
Figures from Ref.~\protect\cite{Berezinsky:2001uv}.
}
\end{figure}

\newpage

\section{Fits of current solar neutrino data}
\label{Fits of current solar neutrino data}
%
%
\begin{description}
\item[Two-Neutrino $\nu_e\to\nu_\mu,\nu_\tau$ Oscillations:]
\protect\cite{Ahmad:2002ka,Barger:2002iv,Bandyopadhyay:2002xj,%
Bahcall:2002hv,Aliani:2002ma,%
deHolanda:2002pp,Fukuda:2002pe,%
Strumia:2002rv,Fogli:2002pt,Fogli:2002pb,Maltoni:2002ni}
\item[Two-Neutrino $\nu_e\to\nu_s$ Oscillations:]
\protect\cite{Bahcall:2002zh,Maltoni:2002ni}
\item[Three-Neutrino Mixing:]
\protect\cite{Fogli:2002pb}
\item[Four-Neutrino Mixing:]
\protect\cite{hep-ph/0207157}
\item[Spin-Flavor Precession:]
\protect\cite{hep-ph/0206193,Barranco:2002te}
\end{description}

\newpage

\section{$\nu_e\to\nu_\mu,\nu_\tau$ allowed regions from Ref.~\protect\cite{Bahcall:2002hv}}
\label{Bahcall:2002hv}
\begin{figure}[H]
\begin{center}
\includegraphics*[bb=-103 158 700 684, width=0.99\textwidth]{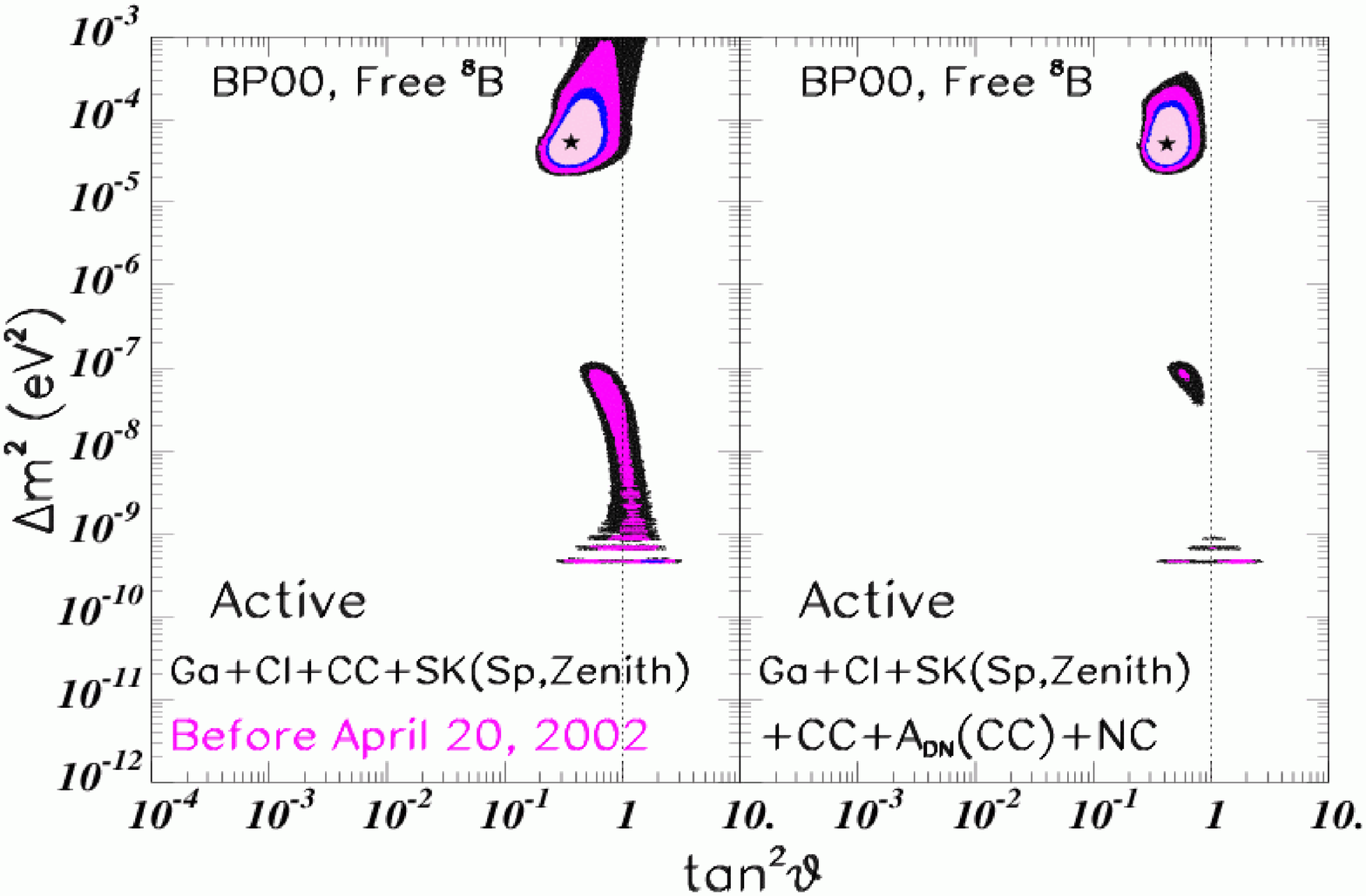}
\end{center}
\caption{ \label{bahcall-0204314-f03}
90\%,
95\%,
99\%,
99.73\% ($3\sigma$)
C.L. regions.
The star marks the best-fit point.
}
\end{figure}

\begin{equation}
\text{Best-fit (LMA):}
\qquad
\tan^2 \vartheta
\simeq
0.42
\,,
\qquad
\Delta{m}^2
\simeq
5.0 \times 10^{-5} \, \mathrm{eV}^2
\label{bahcall-0204314-LMA-best-fit}
\end{equation}
99.73\% C.L. ($3\sigma$) allowed intervals:
\begin{equation}
\begin{array}{lll}
\text{\red{LMA:}}
\qquad
&
0.24 < \tan^2\vartheta < 0.89
\,,
\qquad
&
2.3 \times 10^{-5} < \Delta{m}^2 / \mathrm{eV}^2 < 3.7 \times 10^{-4}
\\
\text{\red{LOW:}}
\qquad
&
0.43 < \tan^2\vartheta < 0.86
\,,
\qquad
&
3.5 \times 10^{-8} < \Delta{m}^2 / \mathrm{eV}^2 < 1.2 \times 10^{-7}
\end{array}
\label{bahcall-0204314-sun-allowed}
\end{equation}

\newpage

\section{$\nu_e\to\nu_\mu,\nu_\tau$ allowed regions from Ref.~\protect\cite{Fogli:2002pt}}
\label{Fogli:2002pt}
\begin{figure}[H]
\begin{center}
\includegraphics*[bb=55 112 534 736, width=0.60\textwidth]{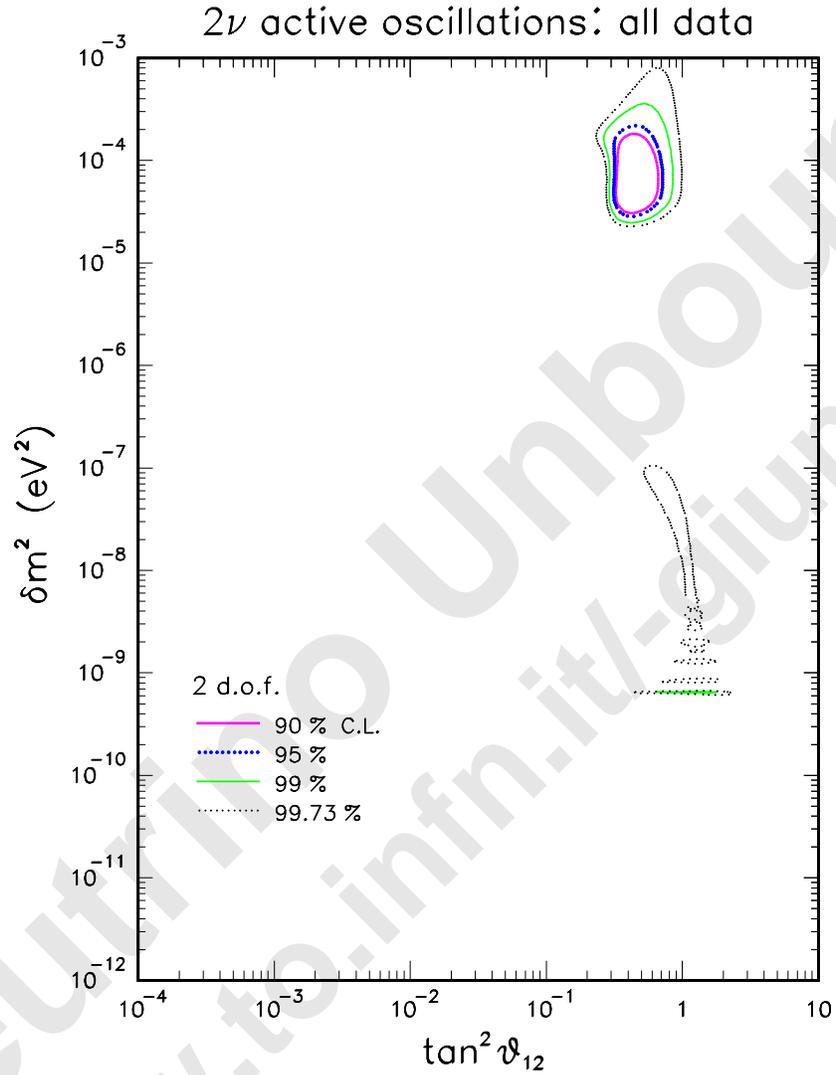}
\end{center}
\caption{ \label{fogli-0206162-f01}
90\%,
95\%,
99\%,
99.73\% ($3\sigma$)
C.L. regions.
}
\end{figure}

\begin{equation}
\text{Best-fit (LMA):}
\qquad
\tan^2 \vartheta
\simeq
0.42
\,,
\qquad
\Delta{m}^2
\simeq
5.5 \times 10^{-5} \, \mathrm{eV}^2
\label{fogli-0206162-LMA-best-fit}
\end{equation}

\newpage

\section{$\nu_e\to\nu_\mu,\nu_\tau$ allowed regions from Ref.~\protect\cite{deHolanda:2002pp}}
\label{deHolanda:2002pp}
\begin{figure}[H]
\begin{center}
\includegraphics*[bb=13 74 513 676, width=0.60\textwidth]{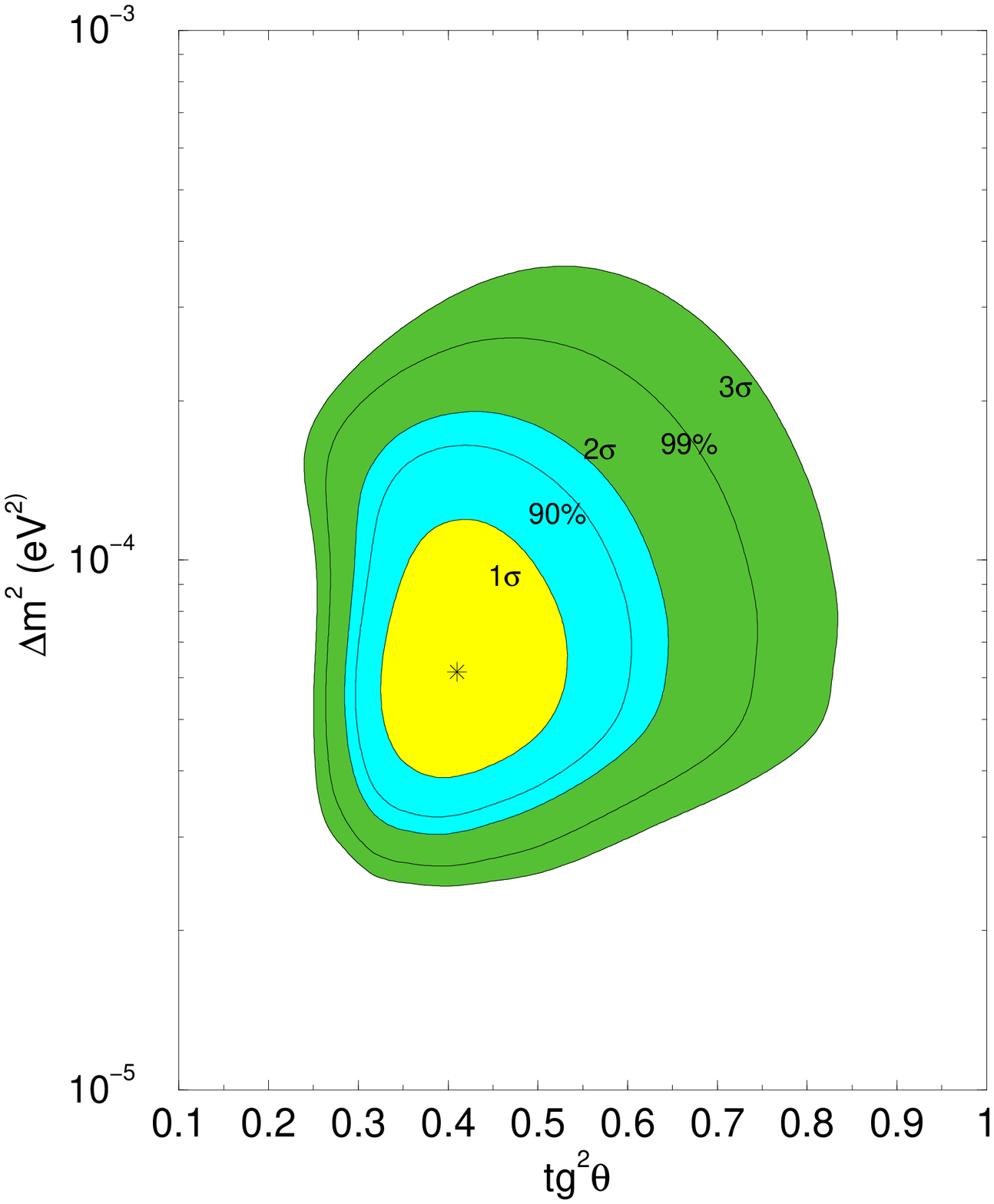}
\end{center}
\caption{ \label{deholanda-0205241-f02}
68.3\% ($1\sigma$)
90\%,
95.5\% ($2\sigma$),
99\%,
99.73\% ($3\sigma$)
C.L. regions.
The star marks the best-fit point.
}
\end{figure}

\begin{equation}
\text{Best-fit (LMA):}
\qquad
\tan^2 \vartheta
\simeq
0.41
\,,
\qquad
\Delta{m}^2
\simeq
6.15 \times 10^{-5} \, \mathrm{eV}^2
\label{deholanda-0205241-LMA-best-fit}
\end{equation}
99.73\% C.L. ($3\sigma$) allowed intervals:
\begin{equation}
\text{\red{LMA:}}
\qquad
0.2 < \tan^2\vartheta < 0.84
\,,
\qquad
2.3 \times 10^{-5} < \Delta{m}^2 / \mathrm{eV}^2 < 3.6 \times 10^{-4}
\label{deholanda-0205241-sun-allowed}
\end{equation}

\newpage

\section{KamLAND $\Longrightarrow$ LMA}
\label{KamLAND -> LMA}
\begin{center}
%
%
Kamioka Liquid scintillator Anti-Neutrino Detector,
long-baseline reactor $\bar\nu_e$ experiment
\\[0.5\semcm]
Kamioka mine (200 km west of Tokyo),
1000 m underground,
2700 m.w.e.
\\[0.5cm]
\begin{tabular}{rl}
average distance from reactors: 180 km
&
\begin{tabular}{l}
6.7\% of flux from one reactor at 88 km
\\
79\% of flux from 26 reactors at 138--214 km
\\
14.3\% of flux from other reactors at $>$295 km
\end{tabular}
\end{tabular}
\\[0.5cm]
1 kt liquid scintillator detector:
$ \bar\nu_{e} + p \to e^{+} + n $,
energy threshold:
$ E_{\mathrm{th}}^{\bar\nu_{e}p} = 1.8 \, \mathrm{MeV} $
\\[0.5cm]
data taking: 4 March -- 6 October 2002, 145.1 days (162 ton yr)
\protect\cite{hep-ex/0212021}
\end{center}
\begin{align}
\null & \quad
\text{expected number of reactor neutrino events (no osc.):}
\null & \quad & \null
N^{\mathrm{KamLAND}}_{\mathrm{expected}} = 86.8 \pm 5.6
\label{NKamLANDexpected}
\\
\null & \quad
\text{expected number of background events:}
\null & \quad & \null
N^{\mathrm{KamLAND}}_{\mathrm{background}} = 0.95 \pm 0.99
\label{NKamLANDbackground}
\\
\null & \quad
\text{observed number of neutrino events:}
\null & \quad & \null
N^{\mathrm{KamLAND}}_{\mathrm{observed}} = 54
\label{NKamLANDobserved}
\end{align}
\begin{equation}
\frac
{N^{\mathrm{KamLAND}}_{\mathrm{observed}}-N^{\mathrm{KamLAND}}_{\mathrm{background}}}
{N^{\mathrm{KamLAND}}_{\mathrm{expected}}}
=
0.611 \pm 0.085 \pm 0.041
\quad
\text{\protect\cite{hep-ex/0212021}}
\label{RKamLAND}
\end{equation}
\begin{center}
99.95\% C.L. evidence of $\bar\nu_e$ disappearance
\end{center}
\begin{figure}[H]
\begin{center}
\includegraphics*[bb=0 0 434 360, height=0.38\textheight]{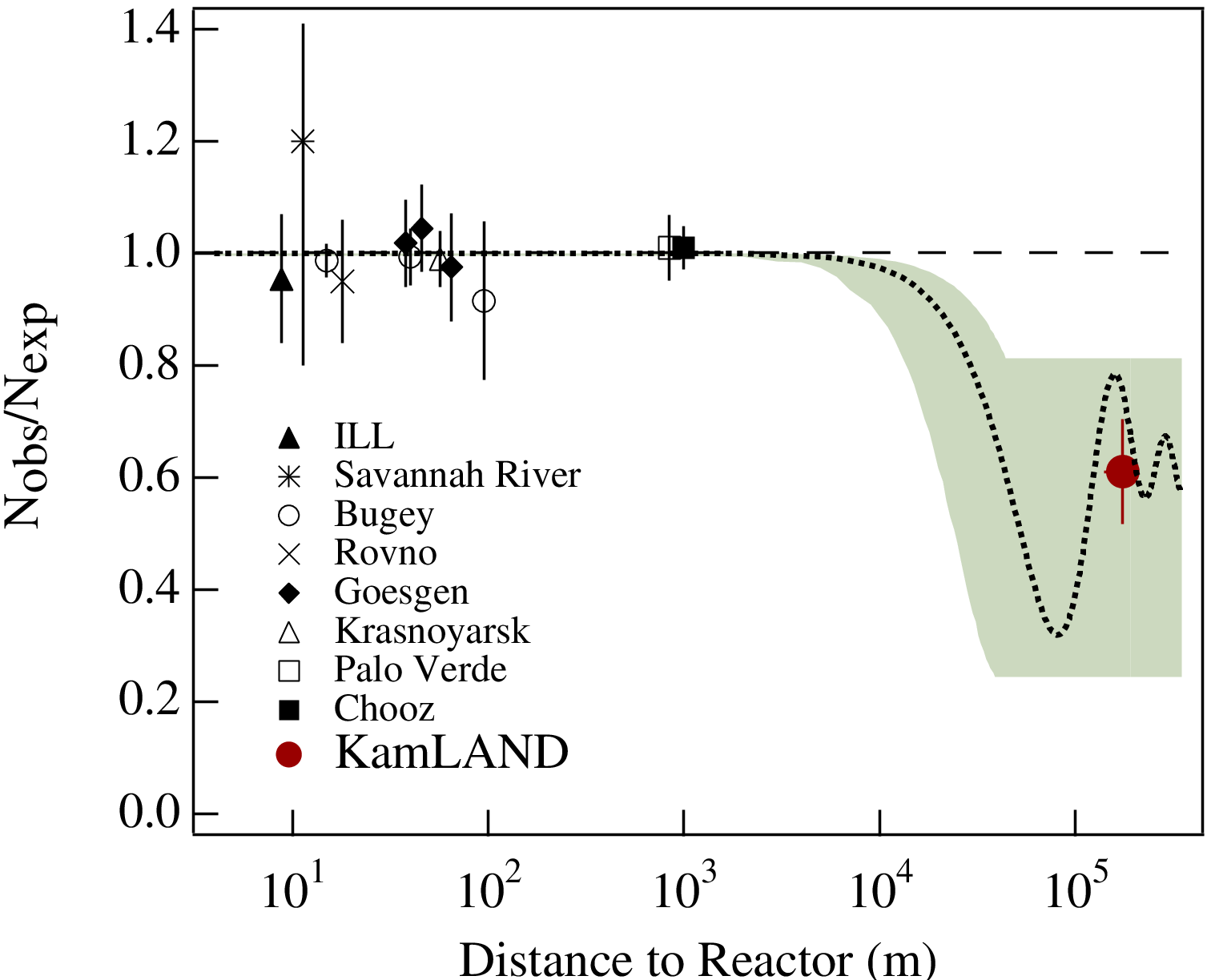}
\end{center}
\caption{ \label{kamland-0212021-f04}
The ratio of measured to expected $\bar \nu_e$ flux from reactor 
experiments.
The shaded region indicates the range of flux 
predictions corresponding to the 95\% C.L. LMA region found in a 
global analysis of the solar neutrino data \protect\cite{Fogli:2002pt}.
The dotted 
curve corresponds to the best-fit values
$\Delta{m}^{2}_{\mathrm{sol}} = 5.5 \times 10^{-5} \, \mathrm{eV}^{2}$
and
$\sin^{2}2\vartheta_{\mathrm{sol}} = 0.83$
found in Ref.~\protect\cite{Fogli:2002pt}.
Figure from Ref.~\protect\cite{hep-ex/0212021}.
}
\end{figure}

\newpage

\begin{figure}[H]
\begin{center}
\subfigure[
Upper panel: Expected reactor $\bar\nu_e$ energy spectrum with contributions
of $\bar\nu_{\mathrm{geo}}$
(antineutrinos emitted by $^{238}$U and $^{232}$Th decays in the earth)
and accidental backround.
Lower panel: Energy spectrum of the observed prompt
events (solid circles with error bars),
along with the expected no oscillation spectrum (upper histogram, with $\bar\nu_{\mathrm{geo}}$
and accidentals shown) and best
fit (lower histogram) including neutrino
oscillations. The shaded band indicates the 
systematic error in the best-fit spectrum.
The vertical dashed line corresponds to the 
analysis threshold at 2.6 MeV.
]{
\includegraphics*[bb=69 50 520 494, width=0.45\textwidth]{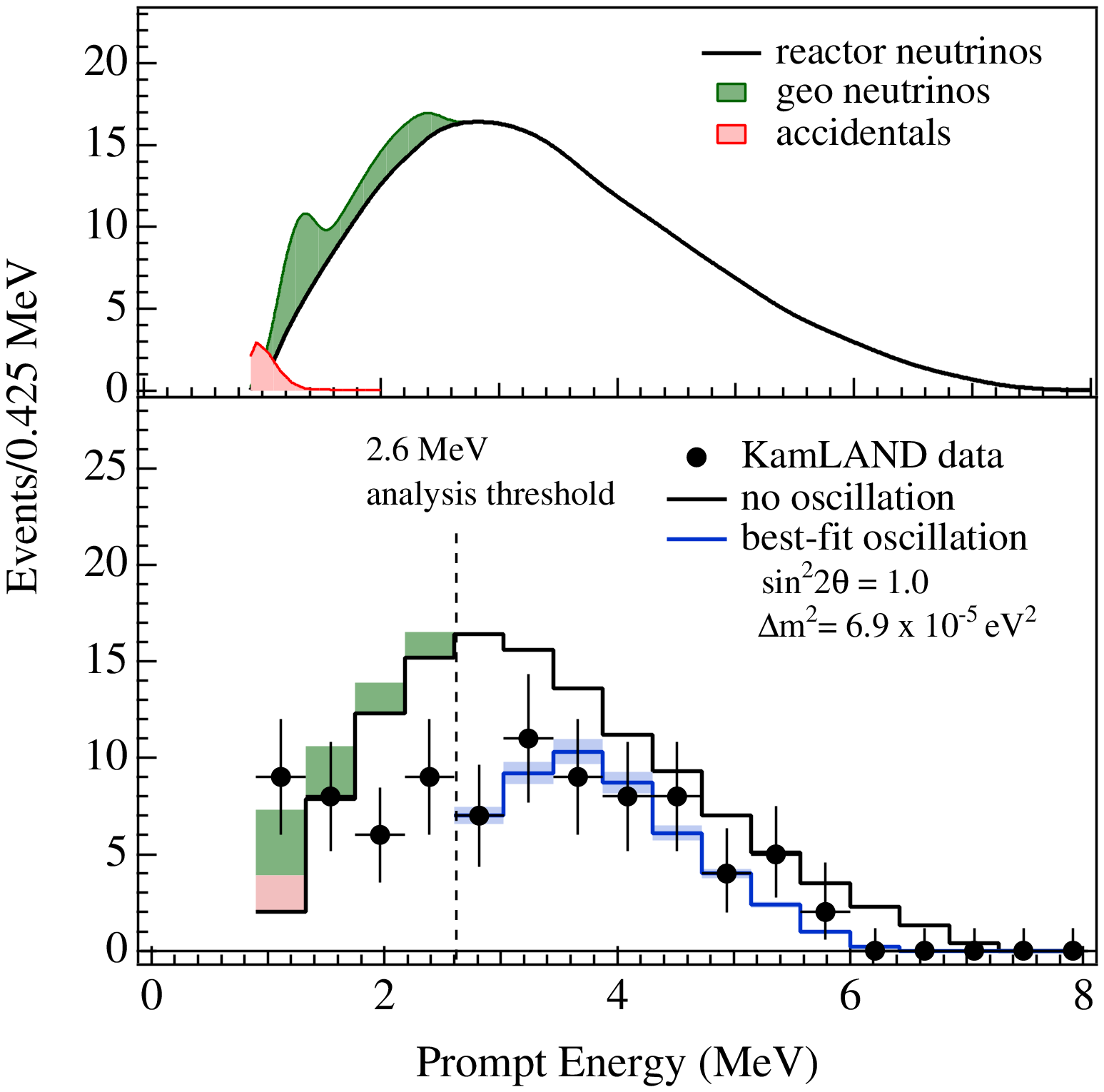}
\label{kamland-0212021-f05}
}
\hfill
\subfigure[
KamLAND
excluded regions of neutrino oscillation parameters
$\Delta{m}^{2}_{\mathrm{KamLAND}} = \mathsf{\Delta{m}^{2}}$
and
$\sin^{2}2 \vartheta_{\mathrm{KamLAND}} = \mathsf{\sin^{2}2\theta}$
for the rate analysis and allowed regions for the combined rate and
energy spectrum analysis at 95\% C.L.
At the top are the 95\% C.L. excluded region from CHOOZ \protect\cite{Apollonio:1999ae,hep-ex/0301017}
and Palo Verde \protect\cite{Boehm:2001ik} experiments, respectively.
The dark area is the 95\% C.L. LMA allowed
region obtained in Ref.~\protect\cite{Fogli:2002pt}.
The thick dot indicates the best fit of
the KamLAND data in Eq.~(\ref{KamLANDbestfit}).
]{
\includegraphics*[bb=0 0 567 539, width=0.45\textwidth]{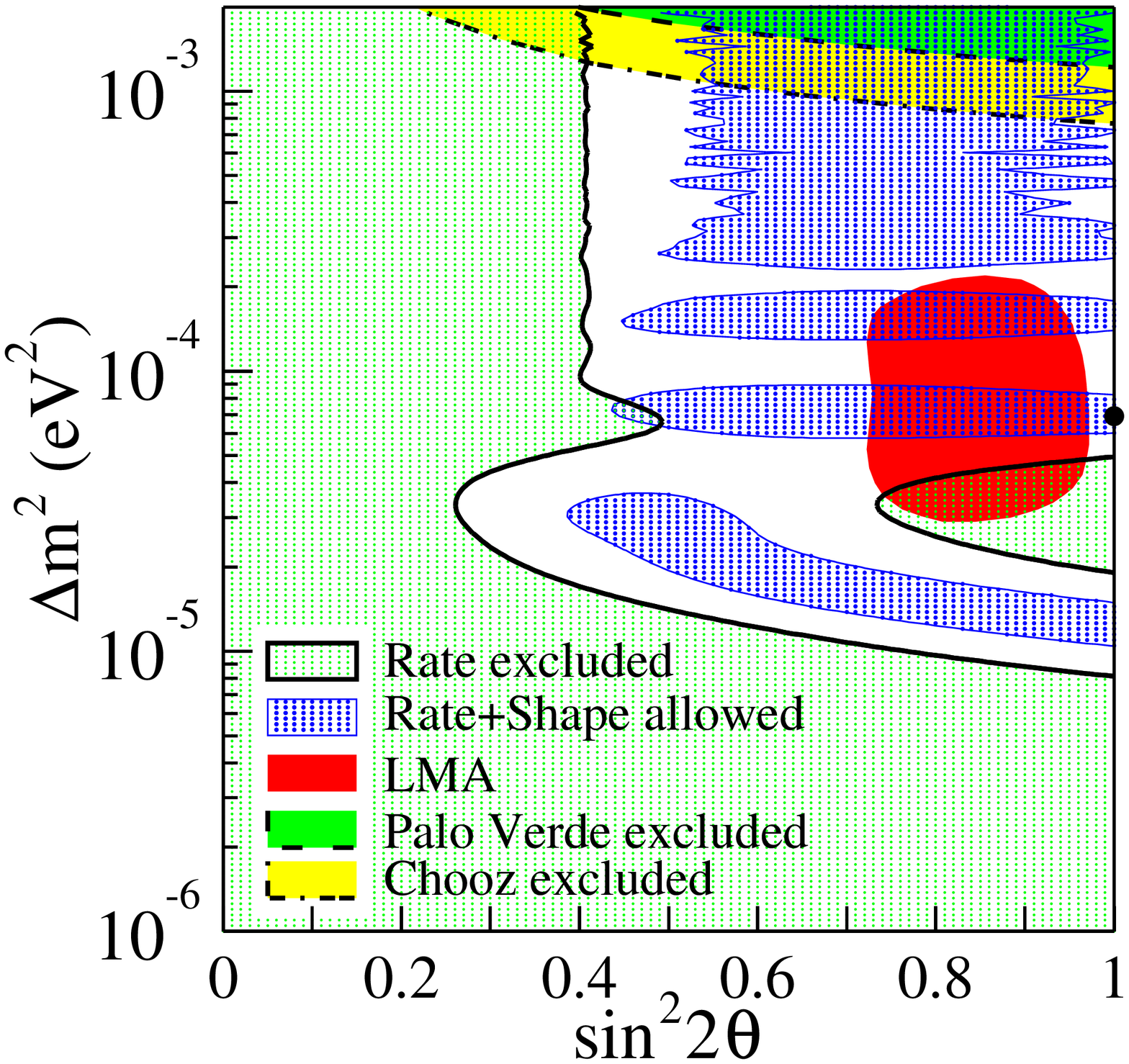}
\label{kamland-0212021-f06}
}
\end{center}
\caption{ \label{kamland-0212021-1}
Figures taken from Ref.~\protect\cite{hep-ex/0212021}.
}
\end{figure}
\begin{equation}
E_{\mathrm{prompt}}
=
E_{\bar\nu_{e}} + m_p - m_n - \overline{T_{n}} + m_e
=
E_{\bar\nu_{e}} - \overline{T_{n}} - 0.8 \, \mathrm{MeV}
\end{equation}
$\overline{T_{n}} = \text{average kinetic energy of the neutron}$;
$m_e$ comes from annihilation of final $e^+$ with $e^-$ in medium
\begin{equation}
\text{best fit:}
\quad
\Delta{m}^{2}_{\mathrm{KamLAND}} = 6.9 \times 10^{-5}\mathrm{eV}^{2}
\,,
\quad
\sin^{2}2\vartheta_{\mathrm{KamLAND}} = 1
\quad
\text{\protect\cite{hep-ex/0212021}}
\label{KamLANDbestfit}
\end{equation}

\vspace{1cm}

\section{Fits of reactor + solar neutrino data}
\label{Fits of reactor + solar neutrino data}
\begin{center}
%
%
%
\protect\cite{hep-ph/0212126,%
hep-ph/0212127,%
hep-ph/0212129,%
hep-ph/0212146,%
hep-ph/0212147,%
hep-ph/0212202,%
hep-ph/0212212,%
hep-ph/0212270,%
hep-ph/0301072}
\end{center}

\newpage

\section{Allowed reactor + solar region from Ref.~\protect\cite{hep-ph/0212127}}
\label{hep-ph/0212127}
\begin{figure}[H]
\begin{center}
\includegraphics*[bb=53 111 530 737, width=0.40\textwidth]{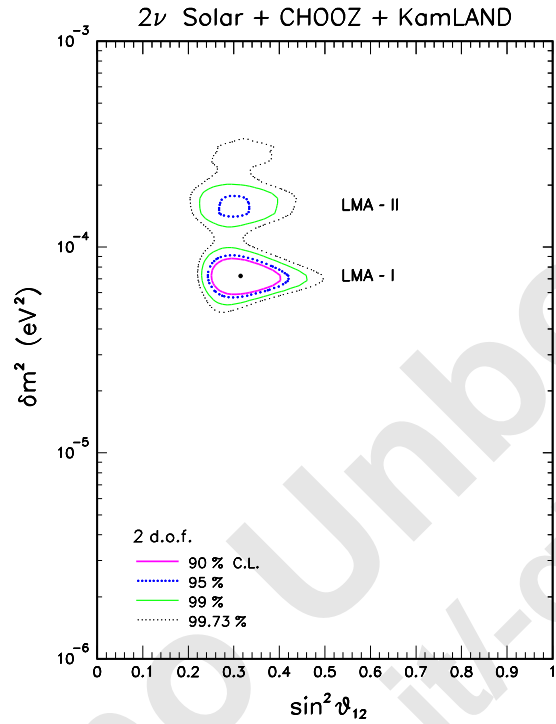}
\end{center}
\caption{ \label{fogli-0212127-f03}
Allowed 90\%, 95\%, 99\%, 99.73\% ($3\sigma$) C.L. regions.
The black dot is the best-fit point.
}
\end{figure}
\begin{equation}
\text{Best-fit:}
\qquad
\sin^2 \vartheta
\simeq
0.315
\,,
\qquad
\Delta{m}^2
\simeq
7.3 \times 10^{-5} \, \mathrm{eV}^2
\label{fogli-0212127-best-fit}
\end{equation}
\begin{figure}[H]
\begin{center}
\includegraphics*[bb=40 242 563 579, width=0.60\textwidth]{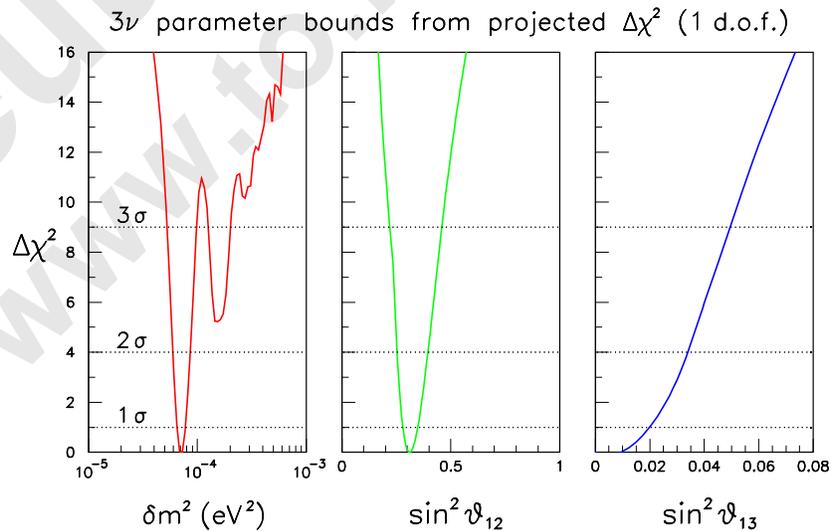}
\end{center}
\caption{ \label{fogli-0212127-f08}
Three-neutrino oscillations: projections of the global $\Delta\chi^2$ function
on the $\delta{m}^2=\Delta{m}^2_{\mathrm{sol}}$, $\sin^2\vartheta_{12}$, $\sin^2\theta_{13}$
axes.
The $n\sigma$ bounds on each parameter
correspond to $\Delta\chi^2=n^2$.
}
\end{figure}

\newpage

\section{Allowed reactor + solar region from Ref.~\protect\cite{hep-ph/0212129}}
\label{hep-ph/0212129}
\begin{figure}[H]
\begin{center}
\includegraphics*[bb=6 21 439 436, width=0.50\textwidth]{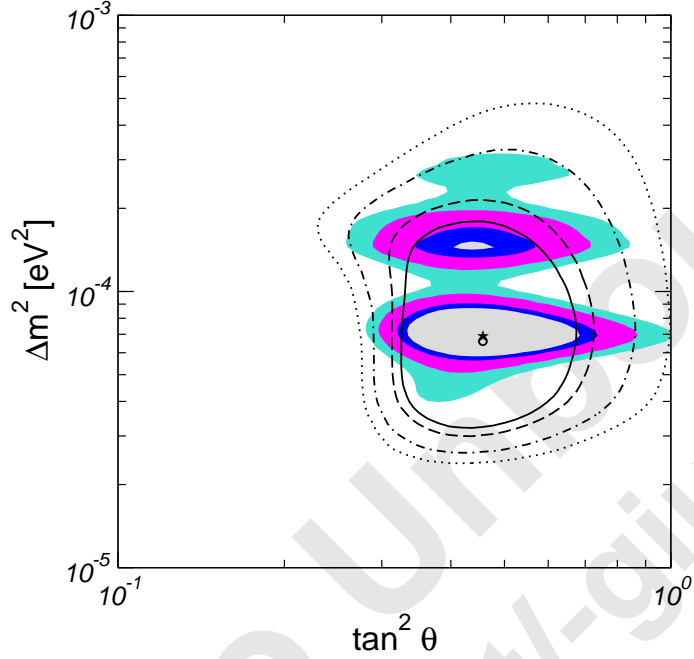}
\end{center}
\caption{ \label{maltoni-0212129-f02}
Allowed 90\%, 95\%, 99\%, 99.73\% ($3\sigma$) C.L. regions.
The hollow lines are the allowed regions from solar and CHOOZ data alone.
The star (dot)
is the best-fit point from the combined
(solar and CHOOZ only) analysis.
}
\end{figure}
\begin{equation}
\text{Best-fit:}
\qquad
\tan^2 \vartheta
\simeq
0.46
\,,
\qquad
\Delta{m}^2
\simeq
6.9 \times 10^{-5} \, \mathrm{eV}^2
\label{maltoni-0212129-bestfit}
\end{equation}
\begin{equation}
\text{99.73\% C.L. ($3\sigma$) allowed interval:}
\quad
0.29 < \tan^2\vartheta < 0.86
\label{maltoni-0212129-t2t}
\end{equation}
\begin{equation}
\text{99.73\% C.L. ($3\sigma$) allowed intervals:}
\quad
\left\{
\begin{array}{l}
5.1 \times 10^{-5}
<
\Delta{m}^2 / \mathrm{eV}^2
<
9.7 \times 10^{-5}
\\
1.2 \times 10^{-4}
<
\Delta{m}^2 / \mathrm{eV}^2
<
1.9 \times 10^{-4}
\end{array}
\right.
\label{maltoni-0212129-dm2}
\end{equation}
\begin{figure}[H]
\begin{center}
\includegraphics*[bb=20 200 689 494, width=0.85\textwidth]{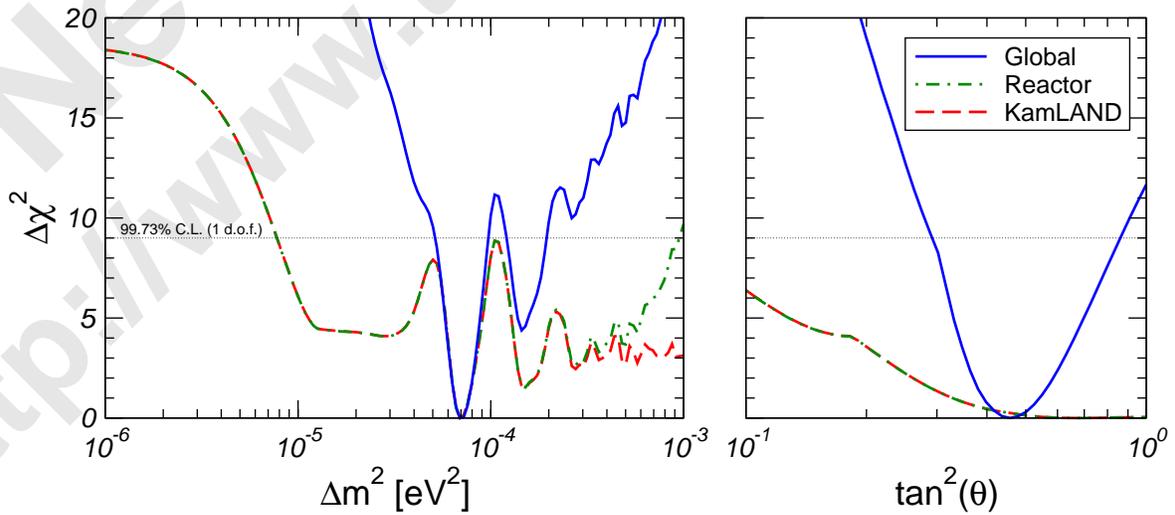}
\end{center}
\caption{ \label{maltoni-0212129-f03}
$\Delta\chi^2$ versus $\Delta{m}^2$ and $\tan^2\vartheta$.
The dashed line
refers to KamLAND alone. The dot-dashed line corresponds
to the full reactor data sample, including both KamLAND and
Chooz. The solid line refers to the global analysis of the
complete solar and reactor data.
}
\end{figure}

\newpage

\section{Allowed reactor + solar region from Ref.~\protect\cite{hep-ph/0212147}}
\label{hep-ph/0212147}
\begin{figure}[H]
\begin{center}
\includegraphics*[bb=104 224 457 594, width=0.50\textwidth]{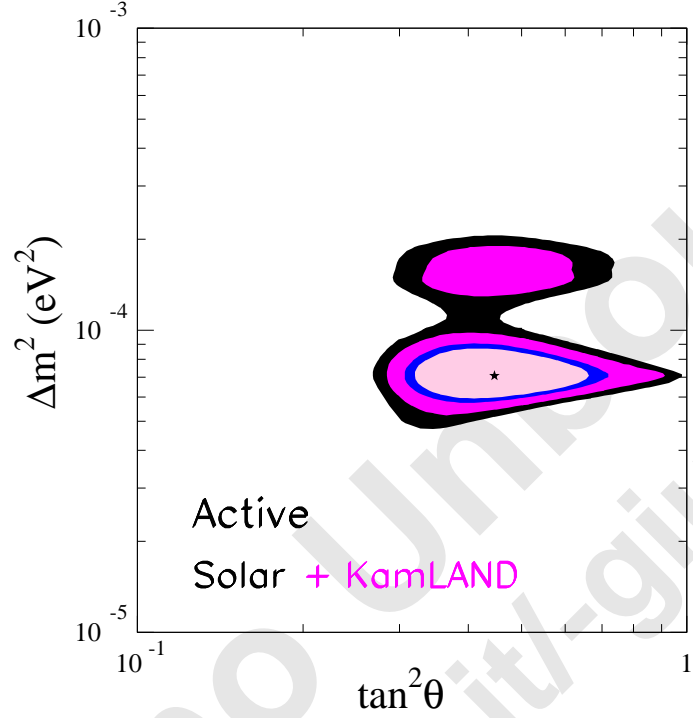}
\end{center}
\caption{ \label{bahcall-0212147-f03}
Allowed 90\%, 95\%, 99\%, 99.73\% ($3\sigma$) C.L. regions.
The global best-fit point is marked by a star.
}
\end{figure}
\begin{equation}
\text{Best-fit:}
\quad
\tan^2 \vartheta
\simeq
0.45
\,,
\quad
\Delta{m}^2
\simeq
7.1 \times 10^{-5} \, \mathrm{eV}^2
\,,
\quad
\frac{\Phi_{^8\mathrm{B}}}{\Phi_{^8\mathrm{B}}^{\mathrm{SSM}}} = 1.00
\label{bahcall-0212147-best-fit}
\end{equation}
\begin{equation}
\text{99.73\% C.L. ($3\sigma$) allowed interval:}
\quad
0.28 < \tan^2\vartheta < 0.91
\label{bahcall-0212147-t2t}
\end{equation}
\begin{equation}
\text{$^8\mathrm{B}$ neutrino flux:}
\quad
\Phi_{^8\mathrm{B}} = 1.00 \pm 0.06 \, \Phi_{^8\mathrm{B}}^{\mathrm{SSM}}
\label{bahcall-0212147-f8b}
\end{equation}
\begin{equation}
\text{sterile neutrino component ($\nu_e \to \cos\eta \nu_a + \sin\eta \nu_s $):}
\quad
\sin^2\eta < 0.52
\,
(3\sigma)
\label{bahcall-0212147-sterile}
\end{equation}
\begin{figure}[H]
\begin{center}
\includegraphics*[bb=18 254 278 550, clip, width=0.30\textwidth]{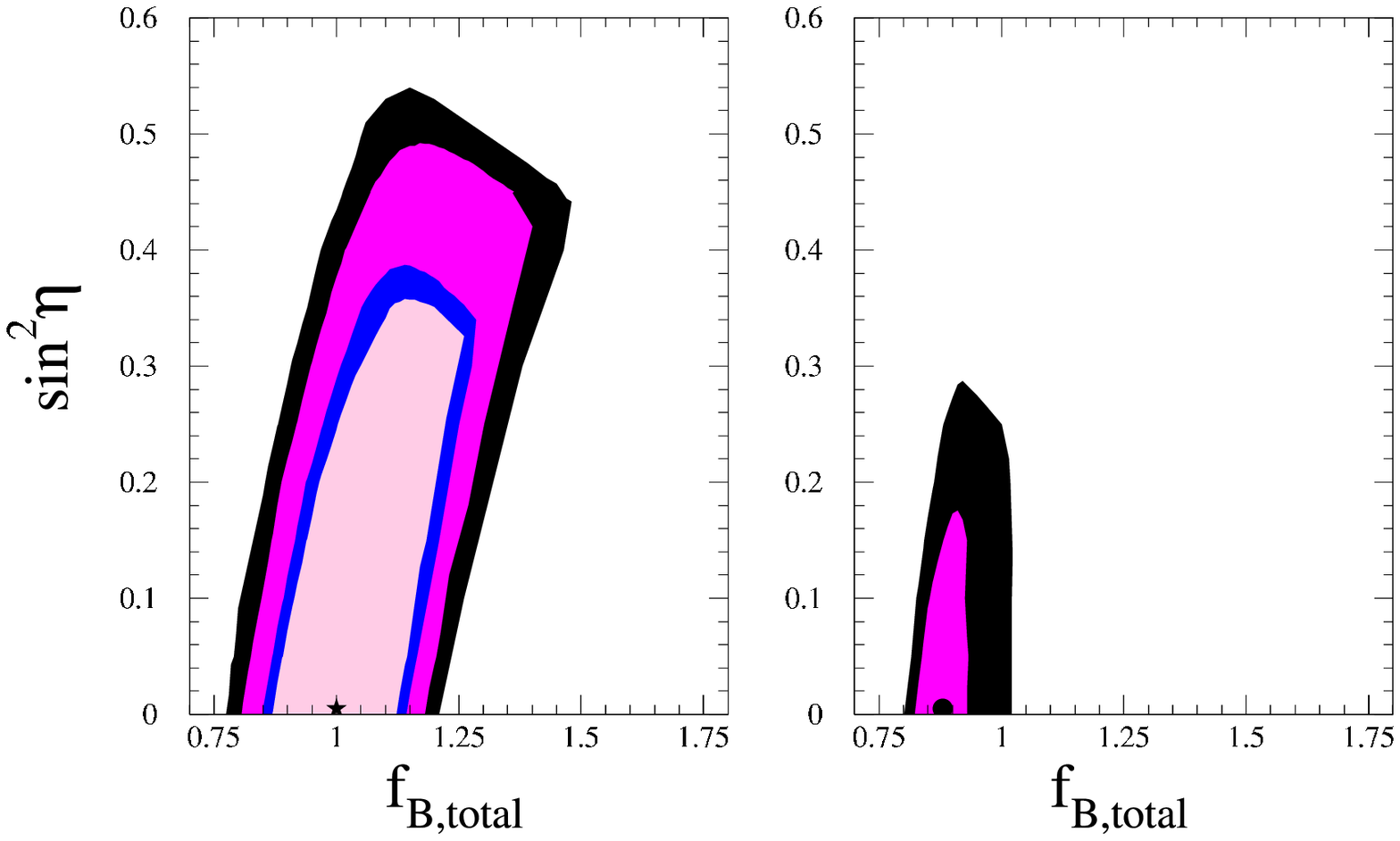}
\end{center}
\caption{ \label{bahcall-0212147-f06}
Allowed 90\%, 95\%, 99\%, 99.73\% ($3\sigma$) C.L. regions
in the $\mathrm{f_{B,total}}$--$\sin^2\eta$ plane,
with $\mathrm{f_{B,total}}=\Phi_{^8\mathrm{B}}/\Phi_{^8\mathrm{B}}^{\mathrm{SSM}}$.
The best-fit point is marked by a star.
}
\end{figure}

\newpage

\section{Allowed reactor + solar region from Ref.~\protect\cite{hep-ph/0212270}}
\label{hep-ph/0212270}
\begin{figure}[H]
\begin{center}
\includegraphics*[bb=22 81 513 676, width=0.60\textwidth]{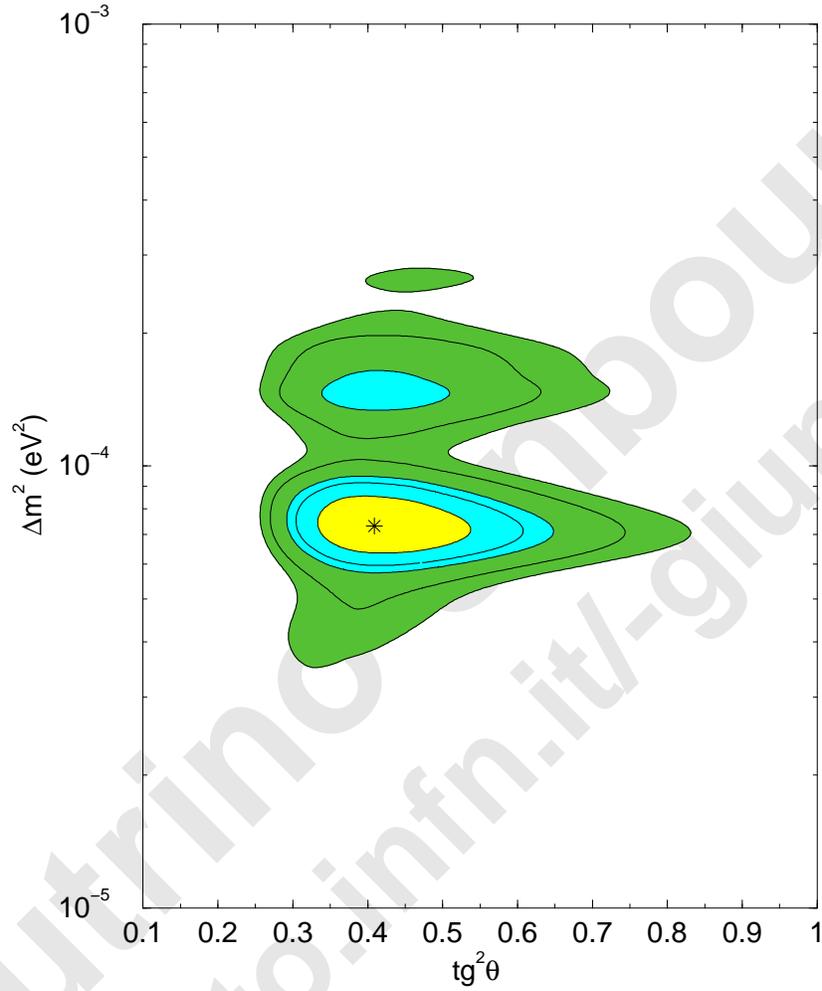}
\end{center}
\caption{ \label{deholanda-0212270-f04}
Allowed
68.3\% ($1\sigma$)
90\%,
95\%,
99\%,
99.73\% ($3\sigma$)
C.L. regions.
The best-fit point is marked by a star.
}
\end{figure}
\begin{equation}
\text{Best-fit:}
\quad
\tan^2 \vartheta
\simeq
0.41
\,,
\quad
\Delta{m}^2
\simeq
7.3 \times 10^{-5} \, \mathrm{eV}^2
\,,
\quad
\frac{\Phi_{^8\mathrm{B}}}{\Phi_{^8\mathrm{B}}^{\mathrm{SSM}}} = 1.05
\label{deholanda-0212270-bestfit}
\end{equation}
\begin{equation}
\text{99.73\% C.L. ($3\sigma$) bounds:}
\quad
\Delta{m}^2 < 2.8 \times 10^{-4} \, \mathrm{eV}^2
\,,
\quad
\tan^2\vartheta < 0.84
\label{deholanda-0212270-bounds}
\end{equation}

\newpage

\end{document}